\documentclass[journal, twoside]{IEEEtran}
\usepackage[T1]{fontenc}
\usepackage[utf8]{inputenc}
\usepackage{color}
\usepackage{bm}
\usepackage{algorithm2e}
\usepackage{amsmath}
\usepackage{amsthm}
\usepackage{amssymb}
\usepackage{graphicx}
\usepackage[bookmarks=true,bookmarksnumbered=true,bookmarksopen=true,bookmarksopenlevel=1,
 breaklinks=false,pdfborder={0 0 0},pdfborderstyle={},backref=false,colorlinks=false]
 {hyperref}
\hypersetup{pdftitle={Your Title},
 pdfauthor={Your Name},
 pdfpagelayout=OneColumn, pdfnewwindow=true, pdfstartview=XYZ, plainpages=false}

\makeatletter
\let\oldforeign@language\foreign@language
\DeclareRobustCommand{\foreign@language}[1]{%
  \lowercase{\oldforeign@language{#1}}}
\theoremstyle{plain}
\newtheorem{thm}{\protect\theoremname}
\theoremstyle{plain}
\newtheorem{prop}[thm]{\protect\propositionname}

\usepackage{cite}
\usepackage{titlesec}
\usepackage[caption=false,font=footnotesize]{subfig}
\usepackage{amsmath, amsfonts, amssymb, graphicx}
\usepackage{hyperref}
\hypersetup{
    colorlinks=true,
    linkcolor=blue,
    filecolor=magenta,      
    urlcolor=blue,
    citecolor=blue,
    }
\newtheorem{remark}{Remark}
\DeclareFontFamily{U}{mathx}{}
\DeclareFontShape{U}{mathx}{m}{n}{<-> mathx10}{}
\DeclareSymbolFont{mathx}{U}{mathx}{m}{n}
\DeclareMathAccent{\widecheck}{0}{mathx}{"71}

\DeclareMathOperator{\trace}{tr}

\DeclareMathOperator*{\minimize}{minimize}
\DeclareMathOperator*{\maximize}{maximize}
\DeclareMathOperator*{\st}{subject~to}

\allowdisplaybreaks
\titlespacing{\section}{0pt}{*0.05}{*0.1}
\titlespacing{\subsection}{0pt}{*0.05}{*0.1}
\ifdefined\showcaptionsetup
 \PassOptionsToPackage{caption=false}{subfig}
\fi
\usepackage{subfig}
\makeatother

\providecommand{\propositionname}{Proposition}
\providecommand{\theoremname}{Theorem}

\begin{document}
\title{Fundamental Trade-Offs in Quantized Hybrid Radar Fusion: A CRB-Rate
Perspective}
\author{Akhileswar~Chowdary,~\IEEEmembership{Graduate Student Member,~IEEE,}
Ahmad~Bazzi, Vaibhav~Kumar, Roberto~Bomfin,~and~Marwa~Chafii~\IEEEmembership{Senior~Member,~IEEE}\thanks{Some parts of this paper were presented at IEEE Global Commun. Conf.
(GLOBECOM), Taipei, Taiwan, December 8-12, 2025, {[}DOI: doi.org/10.48550/arXiv.2601.18539{]}. }\thanks{Akhileswar Chowdary is with the NYU WIRELESS, NYU Tandon School of
Engineering, New York University (NYU), Brooklyn, NY 11201 USA (email:
akhileswar.chowdary@nyu.edu).}\thanks{Ahmad Bazzi and Marwa Chafii are with the Engineering Division, New
York University (NYU) Abu Dhabi, Abu Dhabi, United Arab Emirates,
and the NYU WIRELESS, NYU Tandon School of Engineering, Brooklyn,
NY 11201 USA (e-mail: ahmad.bazzi@nyu.edu; marwa.chafii@nyu.edu).}\thanks{Vaibhav Kumar and Roberto Bomfin are with the Engineering Division,
New York University(NYU) Abu Dhabi, Abu Dhabi, United Arab Emirates
(e-mail: vaibhav.kumar@ieee.org; roberto.bomfin@nyu.edu).}}
\markboth{Submitted to IEEE Transactions on Wireless Communications}{Chowdary \MakeLowercase{\emph{et al.}}: Fundamental Trade-Offs in
Quantized Hybrid Radar Fusion: A CRB-Rate Perspective}
\maketitle
\begin{abstract}
Hybrid radar fusion (HRF), which combines monostatic
and bistatic sensing in a common spectrum, offers enhanced spatial
diversity, but is particularly vulnerable to quantization error effects due to the large power imbalance between the direct and reflected
uplink signals. Although finite-resolution analog-to-digital converters
(ADCs) have been considered in the existing literature on integrated
sensing and communication (ISAC), their role in HRF architectures
has not yet been characterized. This paper develops a finite-resolution
quantized sensing–communication framework for HRF systems by deriving
a Cramér–Rao bound (CRB) and achievable uplink rate. Tight lower bounds
on the Fisher information matrix and the communication rate are obtained,
enabling a tractable characterization of finite-resolution quantized
HRF. The fundamental sensing–communication trade-off is then characterized
through two complementary constrained formulations: CRB minimization
subject to per-user uplink rate requirements, and sum-rate maximization
subject to a CRB constraint, whose solutions trace the CRB–rate trade-offs
in HRF. Numerical results reveal how ADC resolution, dynamic range,
and system configuration jointly shape this boundary and show that
HRF performance can degrade sharply under coarse quantization due to the weak bistatic component, providing design guidelines for selecting
ADC architectures and operating regimes in future HRF-enabled ISAC
systems.
\end{abstract}


\IEEEpeerreviewmaketitle{}

\section{Introduction}
\IEEEPARstart{T}{he} envisioned sixth-generation (6G) wireless systems
are expected to support a broad spectrum of emerging applications,
including immersive experiences, haptics, Industry 4.0, intelligent
transportation, localization, remote healthcare, unmanned aerial vehicles
(UAVs), digital twins, and smart cities~\cite{chafii_twelve_2023}.
These services necessitate a seamless integration between communication
and sensing functionalities~\cite{10437283}, giving rise to the
paradigm of integrated sensing and communication (ISAC). Recognizing
its strategic importance, the International Telecommunication Union
(ITU) has identified ISAC as one of six key 6G use cases~\cite{ITU_recommendation}.
ISAC exploits shared spectrum, waveforms, and hardware to embed radar
sensing within communication systems via dual-functional radar–communication
(DFRC) platforms~\cite{liu_toward_2018,vaibhav_star_ris,bomfinmono24}.
However, this joint use of resources introduces a fundamental trade-off
between sensing and communication performance, inherently limiting
the simultaneous achievability of high estimation accuracy and data
throughput~\cite{10147248}.

A substantial body of work has examined this trade-off in DFRC and
ISAC systems using both \emph{estimation-theoretic} and \emph{information-theoretic}
frameworks. In downlink (DL) multiple-input multiple-output (MIMO)
ISAC, Pareto boundaries between the Cramér–Rao bound (CRB) and achievable
rate have been characterized, and optimal transmit covariance or beamforming
strategies have been developed under power, rate, and sensing constraints~\cite{10217169,10251151,11030586,11215603,11031418,10411942}.
These formulations have been extended to multi-static and cooperative
sensing~\cite{11240213}, reconfigurable intelligent surface (RIS)-aided
ISAC~\cite{11184332}, and scenarios involving arbitrary input distributions
and exact CRB–rate trade-offs~\cite{11161907,11208552}. In orthogonal
frequency-division multiplexing (OFDM)-based ISAC, both single-input
single-output (SISO) and MIMO configurations have been investigated
to characterize CRB–rate or mutual information trade-offs, and to
design corresponding beamforming and power allocation schemes~\cite{11079692,11168825,10938036,11271832,10638744,10901856},
as well as waveform-level designs focused on sidelobe control and
delay–Doppler estimation~\cite{11176952,10901655,11223711}.

Despite this extensive literature, the majority of ISAC studies remain
\emph{downlink-centric}, where sensing is based on echoes of base
station (BS) transmissions. In contrast, uplink (UL) sensing has received
comparatively limited attention. The UL ISAC model in~\cite{10623531}
considers multiple single-antenna user equipments (UEs) transmitting
to a BS that performs sensing. However, it focuses solely on frequency–time
resource allocation and does not account for the impact of multiple
UEs on sensing accuracy. Bistatic and distributed ISAC architectures
have been investigated in~\cite{10678871,11303315}, but \emph{these
approaches do not jointly leverage UL UE transmissions and a DFRC-capable
BS within a unified sensing–communication framework}.

To address this limitation, \emph{hybrid radar fusion} (HRF) was proposed
in~\cite{HRF}, wherein a MIMO DFRC BS performs monostatic sensing
using its DL OFDM waveform while simultaneously exploiting UL transmissions
from UEs and their target reflections. By fusing DL echoes and UL
bistatic reflections, HRF achieves enhanced spatial diversity and
improved target observability compared to monostatic DFRC systems.
However, HRF introduces a new hardware constraint. The direct paths
(DPs) of UE UL signals arriving at the BS are typically several orders
of magnitude stronger than the target-reflected components. As a result,
the \emph{dynamic range} (DR) and \emph{resolution} of the analog-to-digital
converters (ADCs) at the BS become critical design factors. If the
ADCs are unable to resolve weak reflections in the presence of strong
DPs, HRF performance degrades significantly.

This hardware-imposed limitation remains largely unaddressed in the
existing ISAC and HRF literature. In particular, the impact of finite-resolution
ADCs and their DR on the fundamental CRB–rate trade-off in HRF systems
has not been explored. This gap is significant, as HRF performance
is governed not only by waveform and beamforming design, but also
by the receiver front-end’s ability to retain bistatic information
under extreme power disparities.

In this work, we extend the HRF framework, proposed in~\cite{HRF},
by incorporating multi-antenna UEs and finite-resolution ADCs at the
BS. We analyze the effects of ADC resolution and DR on both sensing
accuracy and UL communication performance. Specifically, we adopt
the quantized CRB as the sensing metric and derive Bussgang-based
lower bounds on the Fisher information matrix (FIM) and the achievable
UL rate, thereby enabling tractable analysis of quantized HRF systems.
Using these bounds, we characterize the CRB–rate boundary through
two complementary constrained formulations: (i) CRB minimization under
per-UE UL rate constraints, and (ii) UL sum-rate maximization under
a CRB constraint. By sweeping these constraints, we obtain the Pareto-optimal
sensing–communication trade-off under finite-resolution hardware constraints.

To the best of our knowledge, this work is the first to analyze the
impact of ADC's DR and quantization on HRF systems and to establish
the CRB–rate trade-off in the quantized HRF regime. The main contributions
of this paper are summarized as follows:
\begin{itemize}
\item \textbf{Quantized HRF performance bounds}: We derive novel quantized
CRB for HRF systems with finite-resolution ADCs. Leveraging the Bussgang
theorem, we establish a tractable lower bound on the FIM and the UL
rate that explicitly account for ADC resolution and DR effects.
\item \textbf{CRB–rate trade-off characterization}: Based on the derived
bounds, we formulate two constrained optimization problems that characterize
the CRB–rate boundary in quantized HRF systems, yielding optimal precoding
strategies for both the BS and UEs.
\item \textbf{System-level insights}: We analyze how the number of UEs,
UE and BS antenna counts, number of targets, transmit powers, and
spectral-efficiency constraints influence the CRB–rate trade-off and
determine system sensitivity to ADC resolution and DR.
\item \textbf{Numerical validation}: Extensive simulations illustrate how
finite-resolution and DR limitations affect HRF sensing and communication
performance, offering concrete design guidelines for practical system
deployment.
\end{itemize}

\section{System Model\label{sec:System-Model}}

\begin{figure}[tbh]
\begin{centering}
\includegraphics[width=0.7\columnwidth]{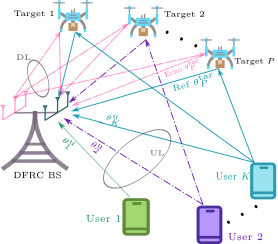}
\par\end{centering}
\raggedright{}\caption{\label{fig:SysMod}HRF system model consisting of a DFRC BS operating
in monostatic mode, receiving echo from $P$ radar targets, and UL
signals from $K$ UEs.}
\end{figure}
We consider a HRF system with a DFRC BS, $K$ communication
UEs, and $P$ point targets in the scene as illustrated in Fig.~\ref{fig:SysMod}.
The DFRC BS is equipped with $N_{\mathrm{t}}^{\mathrm{bs}}$ transmit
antennas and $N_{\mathrm{r}}^{\mathrm{bs}}$ receive antennas arranged
as a uniform linear array (ULA) with inter-element spacing of $\lambda/2$,
where $\lambda$ is the carrier wavelength. Let $\mathcal{K}=\{1,\ldots,K\}$
and $\mathcal{P}=\{1,\ldots,P\}$ denote the index sets of UEs and
targets, respectively. We denote the number of transmit antennas at
UE $k$ as $N_{k}^{\mathrm{u}}$. To address frequency selectivity,
we adopt OFDM over a wideband channel, and assume that the system
operates under a \emph{frequency-division duplex}
(FDD) protocol. The DFRC BS transmits an OFDM waveform in the DL with
$N_{\mathrm{s}}$ OFDM symbols and $N_{\mathrm{c}}$ subcarriers while
simultaneously receiving the echo. Concurrently, the $K$ UEs transmit
their UL signals using OFDM, also with $N_{\mathrm{s}}$ OFDM symbols
and $N_{\mathrm{c}}$ subcarriers.\footnote{Although the system operates under an FDD protocol,
$N_{\mathrm{s}}$ denotes the length of the coherent processing interval
used for joint sensing and communication analysis. The UL and DL occupy
distinct frequency bands and are evaluated over the same number of
OFDM symbols for notational convenience, allowing different UL and
DL symbol lengths does not affect the proposed framework.}

In the UL, the UE subcarrier sets are pairwise disjoint,
i.e., for $i,j\in\mathcal{K}$ and $i\neq j$, $\mathcal{C}_{i}\cap\mathcal{C}_{j}=\varnothing$,
where $\mathcal{C}_{k}$ is the set of subcarrier indices assigned
to UE $k$ and, $\varnothing$ denotes the null set. We define the
set of UL subcarriers indices as $\widehat{\mathcal{N}}\triangleq\bigcup_{k\in\mathcal{K}}\mathcal{C}_{k}$
with $|\widehat{\mathcal{N}}|=N_{\mathrm{c}}$. Let $\widecheck{\mathcal{N}}$
denote the set of DL subcarriers indices used by the DFRC BS with
$|\widecheck{\mathcal{N}}|=N_{\mathrm{c}}$. The subcarrier spacing
in the UL and DL is $\Delta_{\mathrm{f}}=\tfrac{1}{T}$, where $T$
is the (fixed) OFDM symbol duration. For $m\in\widehat{\mathcal{N}}$,
and $m\in\widecheck{\mathcal{N}}$, the subcarrier frequency is $f_{m}=f_{\mathrm{c}}+\left(m+1-\frac{N_{\mathrm{c}}+1}{2}\right)\Delta_{\mathrm{f}}$,
with $f_{\mathrm{c}}$ the carrier frequency. Consequently, the system
bandwidth equals $N_{\mathrm{c}}\Delta_{\mathrm{f}}$ in both DL and
UL. We use the following assumptions in our model:
\begin{itemize}
\item Assumption 1: UEs and the BS are assumed to have
ideal transmit-side hardware. At the BS receiver, each RF chain employs
identical low-resolution ADCs to quantize the in-phase and quadrature
components.
\item Assumption 2: Apart from ADCs, all other components
in the RF chain, e.g., low-noise amplifiers, power amplifiers, and
mixers are ideal.
\item Assumption 3: The sampling rate $f_{\mathrm{s}}$
of the DACs at the UE and the ADCs at the BS is same and the system
is perfectly synchronized.\footnote{The BS employs standardized DL and UL synchronization
procedures during initial access, including synchronization signal
blocks and physical random access channel (PRACH)-based timing advance~\cite{3GPP_TS_38_213}.
These mechanisms jointly compensate for timing and carrier-frequency
offsets, while any residual errors are further mitigated by the cyclic
prefix and receiver-side timing/frequency tracking. As a result, synchronization
errors have a negligible impact on performance, and the assumption
of perfectly synchronized uplink UEs is well justified.}
\item Assumption 4: The inputs to the quantizers across
all the antennas of the BS are assumed to be Gaussian distributed
due to OFDM signaling.\footnote{This assumption holds regardless of the cardinality
of the underlying constellation, as the superposition of a large number
of independent OFDM subcarriers yields approximately Gaussian-distributed
time-domain samples by virtue of the central limit theorem. A detailed
justification is provided in~\cite{Emil_rate_analysis}.}
\end{itemize}
The $i^{\mathrm{th}}$ target of interest is located
at an angle $\theta_{i}^{\mathrm{tar}}$, range $R_{i}^{\mathrm{tar}}$,
and moves with radial velocity $v_{i}^{\mathrm{tar}}$. In what follows,
we introduce the DL transmit signal model at the BS, the UL transmit
signal model at the UE, and the received signal model at the DFRC
BS.

\subsection{Transmit Signal Model at DFRC BS}

For an arbitrary OFDM symbol $\ell$, let $\mathbf{b}_{0}^{(\ell)}=[b_{0,0}^{(\ell)},\ldots,b_{0,N_{\mathrm{c}}-1}^{(\ell)}]^{\mathrm{T}}$
denote the $N_{\mathrm{c}}$ complex data symbols drawn in an independent
and identically distributed (i.i.d) manner from a constellation $\mathcal{S}$
with unit average power, i.e., $\mathbb{E}[\lvert b_{0,m}^{(\ell)}\rvert^{2}]=1$
for $m\in\widecheck{\mathcal{N}}$. The discrete-time (DT) baseband
(BB) DL signal is generated by applying an $N_{\mathrm{c}}$-point
inverse discrete Fourier transform (IDFT) per transmit antenna, which
is given as follows
\begin{equation}
\mathbf{s}_{0}^{(\ell)}[n]=\tfrac{1}{\sqrt{N_{\mathrm{c}}}}\sum\nolimits_{m\in\widecheck{\mathcal{N}}}\mathbf{x}_{m,0}^{(\ell)}e^{j2\pi\frac{mn}{N_{\mathrm{c}}}},\label{eq:tx_symbol_BS}
\end{equation}
where $\mathbf{x}_{m,0}^{(\ell)}\triangleq b_{m,0}^{(\ell)}\mathbf{f}_{m}$
and $\mathbf{f}_{m}\in\mathbb{C}^{N_{\mathrm{t}}^{\mathrm{bs}}\times1}$
is the BS precoder on subcarrier $m$. Appending a cyclic prefix (CP)
of length $N_{\mathrm{CP}}$ yields
\begin{equation}
\mathbf{s}_{0}^{(\ell),\mathrm{CP}}[n]\!=\!\begin{cases}
\mathbf{s}_{0}^{(\ell)}[n\!+\!N_{\mathrm{c}}\!-\!N_{\mathrm{CP}}], & \!\!\!\!n\in\{-N_{\mathrm{CP}},\ldots,-1\}\\
\mathbf{s}_{0}^{(\ell)}[n], & \!\!\!\!n\in\{0,\ldots,N_{\mathrm{c}}-1\}
\end{cases}.\label{eq:tx_symbol_with_CP_BS}
\end{equation}
After passing through DACs, the continuous-time (CT) BB waveform is
given as
\begin{equation}
\mathbf{s}_{0}^{(\ell),\mathrm{CP}}(t)=\sum\nolimits_{n=-N_{\mathrm{CP}}}^{N_{\mathrm{c}}-1}\!\!\!\!\!\mathbf{s}_{0}^{(\ell),\mathrm{CP}}[n]g_{\mathrm{T}}\!\left(t-\ell T_{\mathrm{tot}}-nT_{\mathrm{s}}\right),\label{eq:BB_CT_signal_BS}
\end{equation}
with $g_{\mathrm{T}}(t)$ the transmit pulse and $T_{\mathrm{tot}}\triangleq T+T_{\mathrm{CP}}$
with $T_{\mathrm{CP}}$ the CP duration, and $T_{\mathrm{s}}=\tfrac{T}{N_{\mathrm{c}}}$.
Upconversion to the center frequency $f_{\mathrm{c}}$ produces the
CT passband (PB) signal
\begin{equation}
\mathbf{s}_{0}^{\mathrm{PB}}(t)=\Re\!\big\{\mathbf{s}_{0}^{\mathrm{CP}}(t)e^{j2\pi f_{\mathrm{c}}t}\big\},\label{eq:PB_CT_signal_BS}
\end{equation}
where $\mathbf{s}_{0}^{\mathrm{CP}}(t)\triangleq\sum\nolimits_{\ell=0}^{N_{\mathrm{s}}-1}\mathbf{s}_{0}^{(\ell),\mathrm{CP}}(t).$

\subsection{Transmit Signal Model at UE}

For UE $k\in\mathcal{K}$, $\mathbf{b}_{m,k}^{(\ell)}\in\mathbb{C}^{S_{k}\times1}$
is the vector of complex data symbols whose entries are drawn in an
i.i.d manner from a constellation set $\mathcal{S}_{k}$ with $\mathbb{E}[|b_{m,k,s}^{(\ell)}|^{2}]=1$
for $m\in\widehat{\mathcal{N}}$, $k\in\mathcal{K}$, and $s\in\{0,\ldots,S_{k}-1\}$,
provided $1\leq S_{k}\leq\min\{N_{\mathrm{r}}^{\mathrm{bs}},N_{k}^{\mathrm{u}}\}$.
Note that for a UE $k$, $b_{m,k,s}^{(\ell)}=0$ for $m\notin\mathcal{C}_{k}$.
The DT-BB UL transmit signal is
\begin{equation}
\mathbf{s}_{k}^{(\ell)}[n]=\tfrac{1}{\sqrt{N_{\mathrm{c}}}}\sum\nolimits_{m\in\widehat{\mathcal{N}}}\mathbf{x}_{m,k}^{(\ell)}e^{j2\pi\frac{mn}{N_{\mathrm{c}}}},\label{eq:user_tx_sig}
\end{equation}
where $\mathbf{x}_{m,k}^{(\ell)}\triangleq\mathbf{F}_{m,k}\mathbf{b}_{m,k}^{(\ell)}$
and $\mathbf{F}_{m,k}\in\mathbb{C}^{N_{k}^{\mathrm{u}}\times S_{k}}$
is the $k^{\mathrm{th}}$ UE precoder on subcarrier $m$. After CP
insertion,
\begin{equation}
\mathbf{s}_{k}^{(\ell),\mathrm{CP}}[n]\!=\!\begin{cases}
\mathbf{s}_{k}^{(\ell)}[n\!+\!N_{\mathrm{c}}\!-\!N_{\mathrm{CP}}], & \!\!\!\!n=-N_{\mathrm{CP}},\ldots,-1\\[2pt]
\mathbf{s}_{k}^{(\ell)}[n], & \!\!\!\!n=0,\ldots N_{\mathrm{c}}-1\\[2pt]
\end{cases},\label{eq:UE_symbols}
\end{equation}
and the corresponding CT-BB waveform is
\[
\mathbf{s}_{k}^{(\ell),\mathrm{CP}}(t)=\!\!\!\sum\nolimits_{n=-N_{\mathrm{CP}}}^{N_{\mathrm{c}}-1}\!\!\!\mathbf{s}_{k}^{(\ell),\mathrm{\mathrm{CP}}}[n]g_{\mathrm{T}}\!\left(t-\ell T_{\mathrm{tot}}-nT_{\mathrm{s}}\right).
\]
Upconversion to $f_{\mathrm{c}}$ yields the CT-PB transmit signal
of UE $k$ given by
\begin{equation}
\mathbf{s}_{k}^{\mathrm{PB}}(t)=\Re\!\left\{ \mathbf{s}_{k}^{\mathrm{CP}}(t)e^{j2\pi f_{\mathrm{c}}t}\right\} ,\label{eq:PB_UE_signal}
\end{equation}
where $\mathbf{s}_{k}^{\mathrm{CP}}(t)\triangleq\sum\nolimits_{\ell=0}^{N_{\mathrm{s}}-1}\mathbf{s}_{k}^{(\ell),\mathrm{CP}}(t)$.

\subsection{Received Signal at DFRC BS}

The received signal at the BS consists of two components:
(i) the echo of the BS DL transmission reflected by the $P$ targets,
and (ii) the UL signals transmitted by the $K$ UEs, which include
both the direct UE–BS paths and the UE–target–BS reflected paths.
The received PB signal is then down-converted and passed through a
low-pass filter (LPF) to obtain a CT-BB signal. The BB version of
the received signal is given as follows
\begin{multline}
\mathbf{y}(t)=\left[\mathbf{y}^{\mathrm{PB}}(t)e^{-j2\pi f_{\mathrm{c}}t}\right]\ast g_{\mathrm{R}}(t)=\big[\big\{\mathbf{H}^{\mathrm{echo}}(t)\!\ast\!\mathbf{s}_{0}^{\mathrm{CP}}(t)\big\}\\
+\!\sum\nolimits_{k\in\mathcal{K}}\big\{\mathbf{H}_{k}^{\mathrm{UL}}(t)\!\ast\!\mathbf{s}_{k}^{\mathrm{CP}}(t)\big\}+\mathbf{w}(t)\big]\!\ast g_{\mathrm{R}}(t),\label{eq:y_t}
\end{multline}
where $\mathbf{y}^{\mathrm{PB}}\in\mathbb{C}^{N_{\mathrm{r}}^{\mathrm{bs}}\times1}$
is the received PB signal, $\mathbf{H}^{\mathrm{echo}}(t)\in\mathbb{C}^{N_{\mathrm{r}}^{\mathrm{bs}}\times N_{\mathrm{t}}^{\mathrm{bs}}}$
is the combined PB echo channel (from BS-$P$ targets-BS), $\mathbf{H}_{k}^{\mathrm{UL}}(t)\in\mathbb{C}^{N_{\mathrm{r}}^{\mathrm{bs}}\times N_{k}^{\mathrm{u}}}$
is the PB UL channel between the $k$-th UE and the BS, $\mathbf{w}(t)\in\mathbb{C}^{N_{\mathrm{r}}^{\mathrm{bs}}\times1}$
is the circularly-symmetric complex additive white Gaussian noise
(AWGN), and $g_{\mathrm{R}}(t)$ is the LPF that matches to $g_{\mathrm{T}}(t)$.
The expressions for $\mathbf{H}^{\mathrm{echo}}(t)$ and $\mathbf{H}_{k}^{\mathrm{UL}}(t)$
are, respectively, given as follows
\begin{equation}
\mathbf{H}^{\mathrm{echo}}(t)=\sum_{p\in\mathcal{P}}\tilde{\alpha}_{p}\mathbf{a}_{\mathrm{R}}(\theta_{p}^{\mathrm{tar}})\mathbf{a}_{\mathrm{T}}^{\mathrm{T}}(\theta_{p}^{\mathrm{tar}})e^{j2\pi f_{\mathrm{d},p}^{\mathrm{tar}}t}\delta(t-\tau_{p}^{\mathrm{tar}}),\label{eq:bbecho}
\end{equation}
\begin{equation}
\mathbf{H}_{k}^{\mathrm{UL}}(t)=\tilde{\alpha}_{k}^{\mathrm{u}}\Big(\widehat{\mathbf{H}}_{k}(t)+\widecheck{\mathbf{H}}_{k}(t)\Big)+\breve{\mathbf{H}}_{\ell,k}(t),\label{eq:bbul}
\end{equation}
where $\widehat{\mathbf{H}}_{k}(t)=\sqrt{\xi/(1+\xi)}\mathbf{a}_{\mathrm{R}}(\theta_{k}^{\mathrm{r,u}})\mathbf{a}_{\mathrm{T},k}^{\mathrm{T}}(\theta_{k}^{\mathrm{t,u}})\delta(t-\tau_{k}^{\mathrm{u}})$,
$\widecheck{\mathbf{H}}_{k}$ are the line-of-sight (LoS) and non-line-of-sight
(NLoS) parts of the DP channel in between $k^{\mathrm{th}}$ UE and
the BS, $\breve{\mathbf{H}}_{\ell,k}(t)=\sum\nolimits_{j\in\mathcal{P}_{k}}\tilde{\alpha}_{k,j}^{\mathrm{u}}\mathbf{a}_{\mathrm{R}}(\theta_{j}^{\mathrm{tar}})\mathbf{a}_{\mathrm{T},k}^{\mathrm{T}}(\theta_{k,j}^{\mathrm{u}})e^{j2\pi f_{\mathrm{d},j}^{\mathrm{tar}}t}\delta(t-\tau_{k,j}^{\mathrm{u}})$
is the channel of the reflected paths in between $k$-th UE, targets
that $k$-th UE can see, and the BS, $\xi$ is the Rician factor,
$\tilde{\alpha}_{p}$, $\tilde{\alpha}_{k}^{\mathrm{u}}$, and $\tilde{\alpha}_{k,j}^{\mathrm{u}}$
are the distance dependent path-loss terms including the reflection
coefficients of targets with their expressions given in~\cite[eqns. (5), (7), and (8)]{HRF},
$\tau_{p}^{\mathrm{tar}}=2R_{p}^{\mathrm{tar}}/c$, $f_{\mathrm{d},p}^{\mathrm{tar}}=2f_{\mathrm{c}}v_{p}^{\mathrm{tar}}/c$,
$c$ is the speed of light, $\theta_{k}^{\mathrm{r,u}}$ is the AoA
of the DP from the $k$-th UE at the BS, $\theta_{k}^{\mathrm{t,u}}$
is the AoD of the DP at the $k$-th UE, $\tau_{k}^{\mathrm{u}}$ is
the delay of the DP from the $k$-th UE to the BS, $\theta_{k,j}^{\mathrm{u}}$
is the AoD at the $k$-th UE towards$j^{\mathrm{th}}$ target, $\tau_{k,j}^{\mathrm{u}}$
is the delay of the $k$-th UE-$j$-th target-BS link, $\delta(\cdot)$
is the Dirac delta function, and $\mathcal{P}_{k}$ denotes the set
of targets observed by UE $k$. Note that $\mathcal{P}_{k}\subseteq\mathcal{P}$
$\forall k\in\mathcal{K}$ and $\mathcal{P}_{1}\cup\ldots\cup\mathcal{P}_{K}=\mathcal{P}$.
The NLoS part of the DP channel $\widecheck{\mathbf{H}}_{k}(t)$ is
generated using NYUSIM~\cite{nyusim_tut} and includes $\sqrt{1/(\xi+1)}$.
In addition, $\mathbf{a}_{\mathrm{T}}(\theta)\in\mathbb{C}^{N_{\mathrm{t}}^{\mathrm{bs}}\times1}$
is the transmit steering vector at the BS, $\mathbf{a}_{\mathrm{R}}(\theta)\in\mathbb{C}^{N_{\mathrm{r}}^{\mathrm{bs}}\times1}$
is the receive steering vector at the BS, $\mathbf{a}_{\mathrm{T},k}(\theta)\in\mathbb{C}^{N_{k}^{\mathrm{u}}\times1}$
is the transmit steering vector at the $k^{\mathrm{th}}$ UE. For
a ULA, when the signal BW $\ll f_{\mathrm{c}},$ the expression of
the $i^{\mathrm{th}}$ element of the steering vector is $a_{i}(\theta)=\frac{1}{\sqrt{N}}\exp\left[j2\pi\frac{d}{\lambda}(i-1)\sin(\theta)\right]$,
where $N$ is the number of rows of $\mathbf{a}(\theta)$, $d$ is
the spacing between the antenna elements of the array~\cite{HRF}.

\subsubsection{DT Domain Received Signal Model}

The CT-BB received waveform is uniformly sampled
at times $t_{n}^{(\ell)}=nT_{\mathrm{s}}+\ell T_{\mathrm{tot}}$ for
$n\in\left\{ -N_{\mathrm{CP}},\ldots,N_{\mathrm{c}}-1\right\} $.
For $n\in\{0,\ldots,N_{\mathrm{c}}-1\}$, the $\ell$-th DT symbol
after removing the CP is expressed as
\begin{flalign}
 & \mathbf{y}_{n}^{(\ell)}\approx\sum\nolimits_{d=0}^{N_{\mathrm{c}}-1}\mathbf{H}_{\ell,d}^{\mathrm{echo}}\mathbf{s}_{0}^{(\ell)}\left[(n-d)\mathrm{mod}N_{\mathrm{c}}\right]\nonumber \\
 & \!\!+\!\!\sum\nolimits_{k\in\mathcal{K}}\sum\nolimits_{d=0}^{N_{\mathrm{c}}-1}\!\!\mathbf{H}_{\ell,d,k}^{\mathrm{UL}}\mathbf{s}_{k}^{(\ell)}\left[(n-d)\mathrm{mod}N_{\mathrm{c}}\right]\!\!+\!\!\mathbf{w}_{n}^{(\ell)},\label{eq:tdrxsig}
\end{flalign}
where $\mathbf{H}_{\ell,d}^{\mathrm{echo}}$ and $\mathbf{H}_{\ell,d,k}^{\mathrm{UL}}$
denote the effective discrete-time echo and UL channels after receive
filtering. The corresponding expressions are respectively given as
follows
\begin{equation}
\mathbf{H}_{\ell,d}^{\mathrm{echo}}\!=\!\sum_{p\in\mathcal{P}}\alpha_{p}\mathbf{a}_{\mathrm{R}}(\theta_{p}^{\mathrm{tar}})\mathbf{a}_{\mathrm{T}}^{\mathrm{T}}(\theta_{p}^{\mathrm{tar}})c_{d}(\tau_{p}^{\mathrm{tar}})e^{j2\pi f_{\mathrm{d},p}^{\mathrm{tar}}\ell T_{\mathrm{tot}}},\label{eq:bbecho_2}
\end{equation}
\begin{equation}
\mathbf{H}_{\ell,d,k}^{\mathrm{UL}}\!=\!\alpha_{k}^{\mathrm{u}}\Big(\widehat{\mathbf{H}}_{d,k}+\widecheck{\mathbf{H}}_{d,k}\Big)+\breve{\mathbf{H}}_{\ell,d,k},\label{eq:bbul_2}
\end{equation}
where $\alpha_{p}=\tilde{\alpha}_{p}e^{-j2\pi f_{\mathrm{c}}\tau_{p}^{\mathrm{tar}}}$,
$\alpha_{k}^{\mathrm{u}}=\tilde{\alpha}_{k}^{\mathrm{u}}e^{-j2\pi f_{\mathrm{c}}\tau_{k}^{\mathrm{u}}}$,
$\alpha_{k,j}^{\mathrm{u}}=\tilde{\alpha}_{k,j}^{\mathrm{u}}e^{-j2\pi f_{\mathrm{c}}\tau_{k,j}^{\mathrm{u}}}$,
$\widehat{\mathbf{H}}_{d,k}=\sqrt{\xi/(\xi+1)}\mathbf{a}_{\mathrm{R}}(\theta_{k}^{\mathrm{r,u}})\mathbf{a}_{\mathrm{T},k}^{\mathrm{T}}(\theta_{k}^{\mathrm{t,u}})c_{d}(\tau_{k}^{\mathrm{u}})$,
$\breve{\mathbf{H}}_{\ell,d,k}=\sum\nolimits_{j\in\mathcal{P}_{k}}\alpha_{k,j}^{\mathrm{u}}\mathbf{a}_{\mathrm{R}}(\theta_{j}^{\mathrm{tar}})\mathbf{a}_{\mathrm{T},k}^{\mathrm{T}}(\theta_{k,j}^{\mathrm{u}})c_{d}(\tau_{k,j}^{\mathrm{u}})e^{j2\pi f_{\mathrm{d},j}^{\mathrm{tar}}\ell T_{\mathrm{tot}}}$,
and $c_{d}(\tau)=\int_{-\infty}^{+\infty}g_{\mathrm{T}}(u)g_{\mathrm{R}}^{*}(u-dT_{\mathrm{s}}-\tau)$
for $d\in\mathbb{Z}_{\geq0}$. For a Nyquist sinc pulse, i.e., $g_{\mathrm{T}}(t)=g_{\mathrm{R}}(t)=\mathrm{sinc}\left(t/T_{\mathrm{s}}\right)$,
we have $c_{d}(\tau)=\mathrm{sinc}\left(d-\tau/T_{\mathrm{s}}\right)$
with $\mathrm{sinc}(x)=\frac{\sin{\pi x}}{\pi x}$. To avoid inter-carrier
interference (ICI), $\Delta_{\mathrm{f}}$ is typically chosen to
be greater than the maximum Doppler shift, i.e., for $f_{\mathrm{d}}^{\mathrm{tar}}=\max\{f_{\mathrm{d},i}^{\mathrm{tar}}\}_{i\in\mathcal{P}}$,
we obtain $f_{\mathrm{d}}^{\mathrm{tar}}/\Delta_{\mathrm{f}}\ll1$.
Thus, $f_{\mathrm{d},i}^{\mathrm{tar}}/\Delta_{\mathrm{f}}\ll1$ $\forall$
$i\in\mathcal{P}$. Therefore, the Doppler-induced phase shift, $e^{j2\pi nT_{\mathrm{s}}f_{\mathrm{d},i}^{\mathrm{tar}}}$,
within one symbol duration can be omitted~\cite{mimo_ofdm_isac_lee_swindlehurst}
and the approximation, $e^{j2\pi nT_{\mathrm{s}}f_{\mathrm{d},i}^{\mathrm{tar}}}\approx1$
$\forall$ $i\in\mathcal{P}$ holds.

We can write \eqref{eq:tdrxsig} using block-circulant
matrices, which is given as follow
\begin{multline}
\bar{\mathbf{y}}^{(\ell)}=\bar{\mathbf{H}}_{\ell}^{\mathrm{echo}}\big(\mathbf{F}_{N_{\mathrm{c}}}^{\mathrm{H}}\otimes\mathbf{I}_{N_{\mathrm{t}}^{\mathrm{bs}}}\big)\mathbf{x}_{0}^{(\ell)}\\
+\sum\nolimits_{k\in\mathcal{K}}\bar{\mathbf{H}}_{\ell,k}^{\mathrm{UL}}\big(\mathbf{F}_{N_{\mathrm{c}}}^{\mathrm{H}}\otimes\mathbf{I}_{N_{k}^{\mathrm{u}}}\big)\mathbf{x}_{k}^{(\ell)}+\bar{\mathbf{w}}^{(\ell)},\label{eq:tdrxsig_block2}
\end{multline}
where $\bar{\mathbf{y}}^{(\ell)}\in\mathbb{C}^{N_{\mathrm{r}}^{\mathrm{bs}}N_{\mathrm{c}}\times1}=\big[\big(\mathbf{y}_{0}^{(\ell)}\big)^{\mathrm{T}}\ldots\big(\mathbf{y}_{N_{\mathrm{c}}-1}^{(\ell)}\big)^{\mathrm{T}}\big]^{\mathrm{T}}$,
$\bar{\mathbf{H}}_{\ell}^{\mathrm{echo}}\in\mathbb{C}^{N_{\mathrm{r}}^{\mathrm{bs}}N_{\mathrm{c}}\times N_{\mathrm{t}}^{\mathrm{bs}}N_{\mathrm{c}}}=\mathrm{circ}\big(\mathbf{H}_{\ell,0}^{\mathrm{echo}}\mathbf{H}_{\ell,N_{\mathrm{c}}-1}^{\mathrm{echo}}\ldots\mathbf{H}_{\ell,1}^{\mathrm{echo}}\big)$,
$\bar{\mathbf{H}}_{\ell,k}^{\mathrm{UL}}\in\mathbb{C}^{N_{\mathrm{r}}^{\mathrm{bs}}N_{\mathrm{c}}\times N_{k}^{\mathrm{u}}N_{\mathrm{c}}}=\mathrm{circ}\big(\mathbf{H}_{\ell,0,k}^{\mathrm{UL}}\mathbf{H}_{\ell,N_{\mathrm{c}}-1,k}^{\mathrm{UL}}\ldots\mathbf{H}_{\ell,1,k}^{\mathrm{UL}}\big)$,
and $\bar{\mathbf{w}}^{(\ell)}\in\mathbb{C}^{N_{\mathrm{r}}^{\mathrm{bs}}N_{\mathrm{c}}\times1}=\big[\big(\mathbf{w}_{0}^{(\ell)}\big)^{\mathrm{T}}\ldots\big(\mathbf{w}_{N_{\mathrm{c}}-1}^{(\ell)}\big)^{\mathrm{T}}\big]^{\mathrm{T}}$
with $\mathrm{circ}(\cdot)$ producing a block circulant matrix by
performing a right-circular shift of the elements given in the argument
of the operator. Moreover, $\mathbf{F}_{N_{\mathrm{c}}}$ is an $N_{\mathrm{c}}\times N_{\mathrm{c}}$
DFT matrix, $\mathbf{x}_{0}^{(\ell)}\in\mathbb{C}^{N_{\mathrm{t}}^{\mathrm{bs}}N_{\mathrm{c}}\times1}=\big[\big(\mathbf{x}_{0,0}^{(\ell)}\big)^{\mathrm{T}}\ldots\big(\mathbf{x}_{N_{\mathrm{c}}-1,0}^{(\ell)}\big)^{\mathrm{T}}\big]^{\mathrm{T}}$,
and $\mathbf{x}_{k}^{(\ell)}\in\mathbb{C}^{N_{k}^{\mathrm{u}}N_{\mathrm{c}}\times1}=\big[\big(\mathbf{x}_{0,k}^{(\ell)}\big)^{\mathrm{T}}\ldots\big(\mathbf{x}_{N_{\mathrm{c}}-1,k}^{(\ell)}\big)^{\mathrm{T}}\big]^{\mathrm{T}}$
$\forall$$k\in\mathcal{K}$.

\section{Quantization and Performance Analysis}

In this section, we develop a unified analytical
framework to characterize the impact of finite-resolution analog-to-digital
conversion on both sensing fidelity and UL communication performance
in the HRF architecture. Section~\ref{subsec:quantizer_model} introduces
the adopted scalar, memoryless $b$-bit quantization model and formalizes
the underlying operating assumptions. In Section~\ref{subsec:achievable_sum_rate},
we derive a frequency-domain achievable UL sum-rate expression under
quantization, culminating in~\eqref{eq:rate}. Section~\ref{sec:FIM_lower_bound}
establishes a tractable lower bound on the FIM, as given in~\eqref{eq:FIM_LB_2},
enabling analytical characterization of estimation performance in
the quantized regime. Finally, Section~\ref{subsec:ADC_DR} provides
a detailed discussion of the role of ADC DR and derives a conservative
resolution requirement to ensure the resolvability of both the DP
and target-reflected components.

\subsection{Quantizer Model\label{subsec:quantizer_model}}

We denote a scalar, memoryless $b$ bit quantizer
by $Q(\cdot)$, which provides $N_{\mathrm{q}}=2^{b}$ output levels.
The sets of reconstruction levels and decision thresholds of $Q(\cdot)$
are, respectively, given by $\mathcal{U}=\{u_{0},\ldots,u_{N_{\mathrm{q}}-1}\}$
and $\mathcal{V}=\{v_{0},\ldots,v_{N_{\mathrm{q}}}\}$, with $v_{0}-\infty$
and $v_{N_{\mathrm{q}}}=+\infty$. Thus, for a complex input signal
$x=\Re\{x\}+j\Im\{x\}$, we have $Q(x)=Q(\Re\{x\})+jQ(\Im\{x\})$.
For $\Re\{x\}\in(v_{i},v_{i+1}]$, we have $Q(\Re\{x\})=u_{i}$, where
$i\in\{0,\ldots,N_{\mathrm{q}}-1\}$. $Q(\Im\{x\})$ is obtained analogously.
For vector inputs, $Q(\cdot)$ operates element-wise. For input signals
with standard Gaussian distribution, $\mathcal{U}$ and $\mathcal{V}$
for uniform quantizers are given in~\cite{max_quantizing_1960} and
the optimal $\mathcal{U}$ and $\mathcal{V}$ that achieve MMSE between
the input and output of the quantizer are given by the Lloyd-Max algorithm
in~\cite{max_quantizing_1960}, which in general yields a non-uniform
quantizer. Thus, our derivations are valid for both uniform and non-uniform
quantizers, provided that they are memoryless.\footnote{Extending the analysis to quantizers with memory,
such as sigma-delta ADC would require additional architecture-specific
parameters, such as companding laws, oversampling and noise shaping,
or wrap intervals, which fundamentally alter the signal model and
are therefore beyond the scope of this work.}

\subsection{Achievable Sum Rate\label{subsec:achievable_sum_rate}}

The time-sampled signal is processed through $Q(\cdot)$.
Additionally, we assume the quantizer inputs are Gaussian\footnote{As justified in Footnote 1, the quantizer inputs
are assumed to be Gaussian.}, and optimal quantization is applied based on the
input distribution~\cite{mo_channel_2018}. Leveraging the Bussgang
decomposition, any nonlinear mapping of a Gaussian input can be represented
as a linear term plus a distortion component, which is uncorrelated
with the input~\cite{jjbussgang_1952}. Applying this to the DT received
signal in~\eqref{eq:tdrxsig_block2}, the quantized vector, $\bar{\mathbf{r}}^{(\ell)}=Q(\mathbf{\bar{y}}^{(\ell)})$
is expressed as
\begin{equation}
\bar{\mathbf{r}}^{(\ell)}=\mathbf{B}\bar{\mathbf{y}}^{(\ell)}+\bar{\mathbf{w}}_{\mathrm{q}}^{(\ell)},\label{eq:bdfirsteqn}
\end{equation}
where $\bar{\mathbf{r}}^{(\ell)}\in\mathbb{C}^{N_{\mathrm{r}}^{\mathrm{bs}}N_{\mathrm{c}}\times1}=\big[\big(\mathbf{r}_{0}^{(\ell)}\big)^{\mathrm{T}}\ldots\big(\mathbf{r}_{N_{\mathrm{c}}-1}^{(\ell)}\big)^{\mathrm{T}}\big]^{\mathrm{T}}$,
$\bar{\mathbf{w}}_{\mathrm{q}}^{(\ell)}\in\mathbb{C}^{N_{\mathrm{r}}^{\mathrm{bs}}N_{\mathrm{c}}\times1}=\big[\big(\mathbf{w}_{\mathrm{q},0}^{(\ell)}\big)^{\mathrm{T}}\ldots\big(\mathbf{w}_{\mathrm{q},N_{\mathrm{c}}-1}^{(\ell)}\big)^{\mathrm{T}}\big]^{\mathrm{T}}$,
and $\mathbf{w}_{\mathrm{q},n}^{(\ell)}\in\mathbb{C}^{N_{\mathrm{r}}^{\mathrm{bs}}\times1}$
is the non-Gaussian quantization distortion uncorrelated with $\mathbf{y}_{n}^{(\ell)}$.
The matrix $\mathbf{B}\in\mathbb{C}^{N_{\mathrm{r}}^{\mathrm{bs}}N_{\mathrm{c}}\times N_{\mathrm{r}}^{\mathrm{bs}}N_{\mathrm{c}}}$
is the Bussgang gain. With identical $b$-bit ADCs across all RF chains\footnote{This follows from Assumption 1 mentioned in Section~\ref{sec:System-Model}.},
$\mathbf{B}$ reduces to a scalar $\eta=1-\gamma$, where $\gamma$
denotes the inverse of the signal-to-quantization distortion ratio
(SQR). For Gaussian distributed inputs, distortion factor $\gamma$,
a resolution dependent constant is tabulated in~\cite{max_quantizing_1960}
for uniform and non-uniform quantizers, for different bit resolutions.
Consequently, one can write $\mathbf{r}_{n}^{(\ell)}$ for $n\in\{0,\ldots,N_{\mathrm{c}}-1$\}
as 
\begin{flalign}
 & \mathbf{r}_{n}^{(\ell)}=\eta\mathbf{y}_{n}^{(\ell)}+\mathbf{w}_{\mathrm{q},n}^{(\ell)}=\eta\mathbf{x}_{n}^{(\ell)}+\mathbf{\eta w}_{n}^{(\ell)}+\mathbf{w}_{\mathrm{q},n}^{(\ell)},\label{eq:bdsecondeqn}
\end{flalign}
where 
\begin{multline}
\mathbf{x}_{n}^{(\ell)}=\sum\nolimits_{d=0}^{N_{\mathrm{c}}-1}\mathbf{H}_{\ell,d}^{\mathrm{echo}}\mathbf{s}_{0}^{(\ell)}\left[(n-d)\mathrm{mod}N_{\mathrm{c}}\right]\\
+\sum\nolimits_{k\in\mathcal{K}}\sum\nolimits_{d=0}^{N_{\mathrm{c}}-1}\mathbf{H}_{\ell,d,k}^{\mathrm{UL}}\mathbf{s}_{k}^{(\ell)}\left[(n-d)\mathrm{mod}N_{\mathrm{c}}\right].\label{eq:xnl}
\end{multline}
We perform DFT on both sides of~\eqref{eq:bdsecondeqn}. As a result,
for $m\in\widehat{\mathcal{N}}$, and $m\in\widecheck{\mathcal{N}}$,
we get
\begin{flalign}
\Tilde{\mathbf{r}}_{m}^{(\ell)}= & \eta\Tilde{\mathbf{y}}_{m}^{(\ell)}+\Tilde{\mathbf{w}}_{\mathrm{q},m}^{(\ell)},\label{eq:fdbdseceqn}
\end{flalign}
where $\tilde{\mathbf{r}}_{m}^{(\ell)}=\sum_{n=0}^{N_{\mathrm{c}}-1}\mathbf{r}_{n}^{(\ell)}e^{-j2\pi\frac{mn}{N_{\mathrm{c}}}}$,
$\tilde{\mathbf{y}}_{m}^{(\ell)}=\sum_{n=0}^{N_{\mathrm{c}}-1}\mathbf{y}_{n}^{(\ell)}e^{-j2\pi\frac{mn}{N_{\mathrm{c}}}}$,
and $\tilde{\mathrm{\mathbf{w}}}_{\mathrm{q},m}^{(\ell)}=\sum_{n=0}^{N_{\mathrm{c}}-1}\mathbf{w}_{\mathrm{q},n}^{(\ell)}e^{-j2\pi\frac{mn}{N_{\mathrm{c}}}}$.
Due to the subcarrier orthogonality, $\tilde{\mathbf{y}}_{m}^{(\ell)}$
can be written as follows
\begin{equation}
\tilde{\mathbf{y}}_{m}^{(\ell)}=\begin{cases}
\Tilde{\mathbf{H}}_{\ell,m}^{\mathrm{echo}}\mathbf{x}_{m,0}^{(\ell)}+\Tilde{\mathbf{w}}_{m}^{(\ell)}, & \textrm{for}\,m\in\widecheck{\mathcal{N}}\\
\sum_{k\in\mathcal{K}}\Tilde{\mathbf{H}}_{\ell,m,k}^{\mathrm{UL}}\mathbf{x}_{m,k}^{(\ell)}+\Tilde{\mathbf{w}}_{m}^{(\ell)}, & \textrm{for}\,m\in\widehat{\mathcal{N}}
\end{cases}\label{eq:freq_dom_rx_sig}
\end{equation}
where $\tilde{\mathbf{w}}_{m}^{(\ell)}=\sum_{n=0}^{N_{\mathrm{c}}-1}\mathbf{w}_{n}^{(\ell)}e^{-j2\pi\frac{mn}{N_{\mathrm{c}}}}$,
and the expressions of the echo and UL channels in frequency domain,
$\tilde{\mathbf{H}}_{\ell,m}^{\mathrm{echo}}$ and $\tilde{\mathbf{H}}_{\ell,m,k}^{\mathrm{UL}}$
are obtained by taking $N_{\mathrm{c}}$-point DFT of~\eqref{eq:bbecho_2}
and~\eqref{eq:bbul_2}, respectively. The expressions of $\tilde{\mathbf{H}}_{\ell,m}^{\mathrm{echo}}$
and $\tilde{\mathbf{H}}_{\ell,m,k}^{\mathrm{UL}}$ are, respectively,
given as
\begin{equation}
\tilde{\mathbf{H}}_{\ell,m}^{\mathrm{echo}}\!=\!\sum_{p\in\mathcal{P}}\!\!\alpha_{p}\mathbf{a}_{\mathrm{R}}(\theta_{p}^{\mathrm{tar}})\mathbf{a}_{\mathrm{T}}^{\mathrm{T}}(\theta_{p}^{\mathrm{tar}})e^{-j2\pi m\Delta_{\mathrm{f}}\tau_{p}^{\mathrm{tar}}}e^{j2\pi f_{\mathrm{d},p}^{\mathrm{tar}}\ell T_{\mathrm{tot}}},\label{eq:fd_echo_chan}
\end{equation}
\begin{equation}
\tilde{\mathbf{H}}_{\ell,m,k}^{\mathrm{UL}}\!=\!\alpha_{k}^{\mathrm{u}}\Big(\widetilde{\widehat{\mathbf{H}}}_{m,k}+\widetilde{\widecheck{\mathbf{H}}}_{m,k}\Big)+\widetilde{\breve{\mathbf{H}}}_{\ell,m,k},\label{eq:fd_ul_chan}
\end{equation}
where $\widetilde{\widehat{\mathbf{H}}}_{m,k}=\sqrt{\xi/(\xi+1)}\mathbf{a}_{\mathrm{R}}(\theta_{k}^{\mathrm{r,u}})\mathbf{a}_{\mathrm{T},k}^{\mathrm{T}}(\theta_{k}^{\mathrm{t,u}})e^{-j2\pi m\Delta_{\mathrm{f}}\tau_{k}^{\mathrm{u}}}$
and 
\begin{multline}
\widetilde{\breve{\mathbf{H}}}_{\ell,m,k}=\sum\nolimits_{j\in\mathcal{P}_{k}}\!\!\alpha_{k,j}^{\mathrm{u}}\mathbf{a}_{\mathrm{R}}(\theta_{j}^{\mathrm{tar}})\\
\times\mathbf{a}_{\mathrm{T},k}^{\mathrm{T}}(\theta_{k,j}^{\mathrm{u}})e^{-j2\pi m\Delta_{\mathrm{f}}\tau_{k,j}^{\mathrm{u}}}e^{j2\pi f_{\mathrm{d},j}^{\mathrm{tar}}\ell T_{\mathrm{tot}}}.\label{eq:breve_H}
\end{multline}
 Substituting the expression of $\Tilde{\mathbf{y}}_{m}^{(\ell)}$
from \eqref{eq:freq_dom_rx_sig} in \eqref{eq:fdbdseceqn}, we get
\begin{equation}
\Tilde{\mathbf{r}}_{m}^{(\ell)}=\begin{cases}
\eta\Tilde{\mathbf{H}}_{\ell,m}^{\mathrm{echo}}\mathbf{x}_{m,0}^{(\ell)}+\Tilde{\mathbf{e}}_{m}^{(\ell)}, & \forall m\in\widecheck{\mathcal{N}}\\
\sum_{k\in\mathcal{K}}\eta\Tilde{\mathbf{H}}_{\ell,m,k}^{\mathrm{UL}}\mathbf{x}_{m,k}^{(\ell)}+\Tilde{\mathbf{e}}_{m}^{(\ell)}, & \forall m\in\widehat{\mathcal{N}}
\end{cases},\label{eq:tilde_r}
\end{equation}
where $\Tilde{\mathbf{e}}_{m}^{\ell}=\eta\Tilde{\mathbf{w}}_{m}^{(\ell)}+\Tilde{\mathbf{w}}_{\mathrm{q,}m}^{(\ell)}$.
The effective noise $\Tilde{\mathbf{e}}_{m}^{(\ell)}$ is non-Gaussian
due to $\Tilde{\mathbf{w}}_{\mathrm{q},m}^{(\ell)}$. The covariance
matrix of $\Tilde{\mathbf{e}}_{m}^{(\ell)}$ is given as $\mathbf{R}_{\Tilde{\mathbf{e}}^{\ell}\Tilde{\mathbf{e}}^{\ell}}[m]=\mathbb{E}\big[\Tilde{\mathbf{e}}_{m}^{(\ell)}\big(\Tilde{\mathbf{e}}_{m}^{(\ell)}\big)^{\mathrm{H}}\big]=\mathbf{R}_{\Tilde{\mathbf{w}}_{\mathrm{q}}^{(\ell)}\Tilde{\mathbf{w}}_{\mathrm{q}}^{(\ell)}}[m]+\eta^{2}\sigma^{2}\mathbf{I}_{N_{\mathrm{r}}^{\mathrm{bs}}},$
where $\mathbf{R}_{\tilde{\mathbf{w}}_{\mathrm{q}}^{(\ell)}\tilde{\mathbf{w}}_{\mathrm{q}}^{(\ell)}}[m]=\mathbb{E}\big[\tilde{\mathbf{w}}_{\mathrm{q},m}^{(\ell)}\big(\tilde{\mathbf{w}}_{\mathrm{q},m}^{(\ell)}\big)^{\mathrm{H}}\big]$.

As mentioned in~\cite{uplink_achievable_rate},
when we treat $\Tilde{\mathbf{e}}_{m}^{(\ell)}$ as Gaussian distributed
with the same covariance $\mathbf{R}_{\Tilde{\mathbf{e}}^{\ell}\Tilde{\mathbf{e}}^{\ell}}[m]$,
we obtain a lower bound on the sum rate or an achievable sum-rate.
The expression of the achievable sum rate is given as
\begin{multline}
R_{k}=\tfrac{1}{|\mathcal{C}_{k}|}\sum\nolimits_{m\in\mathcal{C}_{k}}\ln\,\mathrm{det}\big(\mathbf{I}_{N_{\mathrm{r}}^{\mathrm{bs}}}+\eta^{2}\Tilde{\mathbf{H}}_{\ell,m,k}^{\mathrm{UL}}\\
\times\mathbf{F}_{m,k}^{\mathrm{u}}\big(\mathbf{F}_{m,k}^{\mathrm{u}}\big)^{\mathrm{H}}\big(\Tilde{\mathbf{H}}_{\ell,m,k}^{\mathrm{UL}}\big)^{\mathrm{H}}\mathbf{R}_{\tilde{\mathbf{e}}^{(\ell)}\Tilde{\mathbf{e}}^{\ell}}^{-1}[m]\big).\label{eq:rate}
\end{multline}
Evaluating $R_{k}$ in~\eqref{eq:rate} requires $\mathbf{R}_{\Tilde{\mathbf{e}}^{(\ell)}\Tilde{\mathbf{e}}^{(\ell)}}[m]$,
which is challenging to compute exactly due to the unknown covariance
of the quantization distortion. To address this, we derive an approximate
closed-form expression for $\mathbf{R}_{\Tilde{\mathbf{w}}_{\mathrm{q}}^{(\ell)}\Tilde{\mathbf{w}}_{\mathrm{q}}^{(\ell)}}[m]$
in Proposition~\ref{prop:prop_1}.
\begin{prop}
\label{prop:prop_1}$\mathbf{R}_{\Tilde{\mathbf{w}}_{\mathrm{q}}^{(\ell)}\Tilde{\mathbf{w}}_{\mathrm{q}}^{(\ell)}}[m]$
can be approximated as 
\begin{flalign}
\mathbf{R}_{\Tilde{\mathbf{w}}_{\mathrm{q}}^{(\ell)}\Tilde{\mathbf{w}}_{\mathrm{q}}^{(\ell)}}[m]\approx\tfrac{\eta(1-\eta)}{N_{\mathrm{c}}}\sum\nolimits_{m'\in\mathcal{N_{\mathrm{c}}}}\mathrm{diag}\big(\mathbf{R}_{\tilde{\mathbf{y}}^{(\ell)}\tilde{\mathbf{y}}^{(\ell)}}[m']\big)\label{eq:approxcov}
\end{flalign}
\end{prop}
\begin{IEEEproof}
\textcolor{blue}{See Appendix~\ref{sec:prop_1_proof}. }
\end{IEEEproof}
Following~\eqref{eq:approxcov}, $\mathbf{R}_{\Tilde{\mathbf{e}}^{\ell}\Tilde{\mathbf{e}}^{\ell}}[m]$
can be written as 
\begin{align}
 & \mathbf{R}_{\Tilde{\mathbf{e}}^{\ell}\Tilde{\mathbf{e}}^{\ell}}[m]=\mathbf{R}_{\Tilde{\mathbf{e}}^{\ell}\Tilde{\mathbf{e}}^{\ell}}=\mathbf{R}_{\tilde{\mathbf{w}}_{\mathrm{q}}^{(\ell)}\tilde{\mathbf{w}}_{\mathrm{q}}^{(\ell)}}[m]+\eta^{2}\mathbf{R}_{\tilde{\mathbf{w}}^{(\ell)}\tilde{\mathbf{w}}^{(\ell)}}[m],\nonumber \\
=\  & \eta\Big[\tfrac{(1-\eta)}{N_{\mathrm{c}}}\sum\nolimits_{m'\in\mathcal{N_{\mathrm{c}}}}\!\!\!\mathrm{diag}\big(\mathbf{R}_{\tilde{\mathbf{y}}^{(\ell)}\tilde{\mathbf{y}}^{(\ell)}}[m']\big)\!\!+\!\eta\sigma^{2}\mathbf{I}_{N_{\mathrm{r}}^{\mathrm{bs}}}\Big],\label{eq:freq_dom_eff_quant_noise_cov}
\end{align}
where 
\begin{equation}
\!\!\!\!\mathbf{R}_{\mathrm{\tilde{\mathbf{y}}}^{(\ell)}\tilde{\mathbf{y}}^{(\ell)}}[m']\!\!=\!\!\begin{cases}
\!\tilde{\mathbf{H}}_{\ell,m'}^{\mathrm{echo}}\mathbf{f}_{m'}\mathbf{f}_{m'}^{\mathrm{H}}\big(\tilde{\mathbf{H}}_{\ell,m'}^{\mathrm{echo}}\big)^{\mathrm{H}}\\
+\sigma^{2}\mathbf{I}_{N_{\mathrm{r}}^{\mathrm{bs}}}, & \!\!\!\!\!\!\!\forall m'\in\mathcal{N}_{\mathrm{DL}}\\
\!\sum\nolimits_{k\in\mathcal{\mathcal{K}}}\!\tilde{\mathbf{H}}_{\ell,m',k}^{\mathrm{UL}}\mathbf{F}_{m',k}\mathbf{F}_{m',k}^{\mathrm{H}}\\
\times\big(\tilde{\mathbf{H}}_{\ell,m',k}^{\mathrm{UL}}\big)^{\mathrm{H}}+\sigma^{2}\mathbf{I}_{N_{\mathrm{r}}^{\mathrm{bs}}}, & \!\!\!\!\!\!\!\forall m'\in\mathcal{N}_{\mathrm{UL}}
\end{cases}.\label{eq:freq_dom_rx_sig_cov_2}
\end{equation}

\subsection{Lower Bound on the FIM of HRF\label{sec:FIM_lower_bound}}

In this section, we derive a lower bound on the FIM
by leveraging the Bussgang theorem. Specifically, we use the result
in~\cite{stein_lower_2014}, which establishes that for a fixed covariance,
among all additive noises, Gaussian noise minimizes the FIM. Therefore,
we obtain a conservative bound by modeling the effective post-quantization
noise as Gaussian. The resulting lower bound is derived using the
frequency-domain representation of the quantized received signal obtained
via the Bussgang decomposition. For each subcarrier, $m\in\widecheck{\mathcal{N}}$
and $\forall m\in\widehat{\mathcal{N}}$, the quantized received signal
can be written as
\begin{flalign}
\Tilde{\mathbf{r}}_{m}^{(\ell)}= & \eta\tilde{\mathbf{x}}_{m}^{(\ell)}+\tilde{\mathbf{e}}_{m}^{(\ell)},\label{eq:tilde_r_m}
\end{flalign}
where $\tilde{\mathbf{x}}_{m}^{(\ell)}=\Tilde{\mathbf{H}}_{\ell,m}^{\mathrm{echo}}\mathbf{x}_{m,0}^{(\ell)}+\sum_{k\in\mathcal{K}}\Tilde{\mathbf{H}}_{\ell,m,k}^{\mathrm{UL}}\mathbf{x}_{m,k}^{(\ell)}$.
For a Gaussian distributed $\tilde{\mathbf{e}}_{m}^{(\ell)}$, the
FIM is given by~\eqref{eq:FIM_LB_1}, shown on the next page, and
CRB is given by~\cite[Sec. 3.9]{Steven_kay_CRB}
\begin{figure*}[tbh]
\begin{multline}
\!\!\!\!\big[\bar{\mathbf{G}}_{\bm{\psi}}\big]_{i,j}=\sum\nolimits_{\ell=0}^{N_{\mathrm{s}}-1}\Big[\sum\nolimits_{m\in\widecheck{\mathcal{N}}}\Big\{2\eta^{2}\Re\Big\{\Big(\tfrac{\partial\tilde{\mathbf{x}}_{m}^{(\ell)}}{\partial\psi_{i}}\Big)^{\mathrm{H}}\mathbf{R}_{\tilde{\mathbf{e}}^{(\ell)}\tilde{\mathbf{e}}^{(\ell)}}^{-1}[m]\tfrac{\partial\tilde{\mathbf{x}}_{m}^{(\ell)}}{\partial\psi_{j}}\Big\}+\mathrm{tr}\Big(\mathbf{R}_{\tilde{\mathbf{e}}^{(\ell)}\tilde{\mathbf{e}}^{(\ell)}}^{-1}[m]\tfrac{\partial\mathbf{R}_{\tilde{\mathbf{e}}^{(\ell)}\tilde{\mathbf{e}}^{(\ell)}}^{-1}[m]}{\partial\psi_{i}}\mathbf{R}_{\tilde{\mathbf{e}}^{(\ell)}\tilde{\mathbf{e}}^{(\ell)}}^{-1}[m] \tfrac{\partial\mathbf{R}_{\tilde{\mathbf{e}}^{(\ell)}\tilde{\mathbf{e}}^{(\ell)}}^{-1}[m]}{\partial\psi_{j}}\Big)\Big\}\\
+\sum\nolimits_{m\in\widehat{\mathcal{N}}}\Big\{2\eta^{2}\Re\Big\{\Big(\tfrac{\partial\tilde{\mathbf{x}}_{m}^{(\ell)}}{\partial\psi_{i}}\Big)^{\mathrm{H}}\mathbf{R}_{\tilde{\mathbf{e}}^{(\ell)}\tilde{\mathbf{e}}^{(\ell)}}^{-1}[m]\tfrac{\partial\tilde{\mathbf{x}}_{m}^{(\ell)}}{\partial\psi_{j}}\Big\}+\mathrm{tr}\Big(\mathbf{R}_{\tilde{\mathbf{e}}^{(\ell)}\tilde{\mathbf{e}}^{(\ell)}}^{-1}[m]\tfrac{\partial\mathbf{R}_{\tilde{\mathbf{e}}_{m}^{(\ell)}\tilde{\mathbf{e}}_{m}^{(\ell)}}^{-1}}{\partial\psi_{i}}\mathbf{R}_{\tilde{\mathbf{e}}^{(\ell)}\tilde{\mathbf{e}}^{(\ell)}}^{-1}[m]\tfrac{\partial\mathbf{R}_{\tilde{\mathbf{e}}^{(\ell)}\tilde{\mathbf{e}}^{(\ell)}}^{-1}[m]}{\partial\psi_{j}}\Big)\Big\}\Big].\label{eq:FIM_LB_1}
\end{multline}
\end{figure*}
\begin{equation}
\big[\bar{\mathbf{C}}_{\bm{\psi}}\big]_{i,i}=\big[\big(\bar{\mathbf{G}}_{\bm{\psi}}\big)^{-1}\big]_{i,i}.\label{eq:CRB_UB}
\end{equation}
Under the low per-antenna SNR assumption, which is valid in practice
for mmWave systems~\cite{mazher_improved_2021}, $\mathbf{R}_{\tilde{\mathbf{y}}^{(\ell)}\tilde{\mathbf{y}}^{(\ell)}}[m']\approx\mathbf{R}_{\tilde{\mathbf{w}}^{(\ell)}\tilde{\mathbf{w}}^{(\ell)}}[m']=\sigma^{2}\mathbf{I}_{N_{\mathrm{r}}^{\mathrm{bs}}},\:\forall m'\in\widecheck{\mathcal{N}}$
and $\forall m'\in\widehat{\mathcal{N}}$. Thus, substituting $\mathbf{R}_{\tilde{\mathbf{y}}^{(\ell)}\tilde{\mathbf{y}}^{(\ell)}}[m']=\sigma^{2}\mathbf{I}_{N_{\mathrm{r}}^{\mathrm{bs}}}$
in \eqref{eq:freq_dom_eff_quant_noise_cov}, the covariance of $\tilde{\mathbf{e}}_{m}^{(\ell)}$
under low per-antenna assumption can be written as
\begin{flalign}
 & \widehat{\mathbf{R}}_{\tilde{\mathbf{e}}^{(\ell)}\tilde{\mathbf{e}}^{(\ell)}}[m]=\widehat{\mathbf{R}}_{\tilde{\mathbf{e}}^{(\ell)}\tilde{\mathbf{e}}^{(\ell)}}\nonumber \\
=\  & \eta\Big[\tfrac{(1-\eta)}{N_{\mathrm{c}}}\!\!\sum_{m'\in\mathcal{N_{\mathrm{c}}}}\!\!\mathrm{diag}\big(\sigma^{2}\mathbf{I}_{N_{\mathrm{r}}^{\mathrm{bs}}}\big)\!+\!\eta\sigma^{2}\mathbf{I}_{N_{\mathrm{r}}^{\mathrm{bs}}}\Big]\!\!=\eta\sigma^{2}\mathbf{I}_{N_{\mathrm{r}}^{\mathrm{bs}}}.\!\!\!\!\label{eq:freq_dom_eff_quant_noise_cov_low_snr}
\end{flalign}
From \eqref{eq:freq_dom_eff_quant_noise_cov_low_snr}, it is clear
that with the low per-antenna SNR assumption, $\mathbf{R}_{\tilde{\mathbf{e}}^{(\ell)}\tilde{\mathbf{e}}^{(\ell)}}[m]$
becomes independent of all parameters of interest. Consequently, $\frac{\partial\mathbf{R}_{\tilde{\mathbf{e}}^{(\ell)}\tilde{\mathbf{e}}^{(\ell)}}^{-1}[m]}{\partial\psi_{i}}=0$
and $\frac{\partial\mathbf{R}_{\tilde{\mathbf{e}}^{(\ell)}\tilde{\mathbf{e}}^{(\ell)}}^{-1}[m]}{\partial\psi_{j}}=0$
in \eqref{eq:FIM_LB_1}. Therefore, a lower bound of the FIM is given
by~\eqref{eq:FIM_LB_2}, shown on the next page, and its corresponding
CRB is given by
\begin{figure*}[tbh]
\begin{equation}
\big[\widehat{\mathbf{G}}_{\bm{\psi}}\big]_{i,j}=\sum\nolimits_{\ell=0}^{N_{\mathrm{s}}-1}\Big[2\eta^{2}\Re\Big\{\sum\nolimits_{m\in\widecheck{\mathcal{N}}}\Big(\tfrac{\partial\tilde{\mathbf{x}}_{m}^{(\ell)}}{\partial\psi_{i}}\Big)^{\mathrm{H}}\widehat{\mathbf{R}}_{\tilde{\mathbf{e}}^{(\ell)}\tilde{\mathbf{e}}^{(\ell)}}^{-1}[m]\tfrac{\partial\tilde{\mathbf{x}}_{m}^{(\ell)}}{\partial\psi_{j}}+\sum\nolimits_{m\in\widehat{\mathcal{N}}}\Big(\tfrac{\partial\tilde{\mathbf{x}}_{m}^{(\ell)}}{\partial\psi_{i}}\Big)^{\mathrm{H}}\widehat{\mathbf{R}}_{\tilde{\mathbf{e}}^{(\ell)}\tilde{\mathbf{e}}^{(\ell)}}^{-1}[m]\tfrac{\partial\tilde{\mathbf{x}}_{m}^{(\ell)}}{\partial\psi_{j}}\Big\}\Big].\label{eq:FIM_LB_2}
\end{equation}
\rule{1\linewidth}{1pt}
\end{figure*}
\begin{equation}
\big[\widehat{\mathbf{C}}_{\bm{\psi}}\big]_{i,i}=\big[\big(\widehat{\mathbf{G}}_{\bm{\psi}}\big)^{-1}\big]_{i,i}.\label{eq:CRB_UB_LS}
\end{equation}

\subsection{ADC Dynamic Range\label{subsec:ADC_DR}}

This section characterizes the role of ADC DR in
resolving strong and weak signal components at the receiver. In finite-resolution
ADCs, setting the input range to avoid clipping of the strongest component
may cause weaker components to fall below the quantization step and
become indistinguishable from quantization noise. This effect is particularly
critical in HRF, where the UL signals received at the BS comprise
strong DP components and much weaker target-reflected components.
If the reflections are not resolvable after quantization, they cannot
be exploited for sensing, which motivates a minimum ADC resolution
(equivalently DR) requirement to ensure resolvability of all reflections.

Consider a $b$-bit uniform ADC with input range
$[-A,A]$, where $A>0$ denotes the maximum admissible input amplitude
and $\mathrm{FSR}=2A$ is the full-scale range. The quantization step
size or the least significant bit (LSB) is $\Delta=\mathrm{FSR}/2^{b}$.
Suppose the ADC input is a sinusoid with peak amplitude $A_{\mathrm{in}}\leq A$,
and define the input back-off factor as $\varpi\triangleq A_{\mathrm{in}}/A\in(0,1]$.
Accordingly, the quantization-limited DR is defined as follows~\cite[Sec. III-A-1]{ADC_DR}
\begin{equation}
\mathrm{DR}=\frac{A_{\mathrm{in}}^{2}/2}{\Delta^{2}/12}=\frac{3}{2}2^{2b}\varpi^{2},
\end{equation}
and in dB scale it becomes
\begin{align}
\mathrm{DR}_{\mathrm{dB}} & =10\log_{10}\left(\frac{3}{2}\right)+20b\log_{10}(2)+20\log_{10}(\varpi)\\
 & \approx6.02b+1.76+20\log_{10}(\varpi).\label{eq:DR_dB}
\end{align}
This expression shows that operating the ADC with an input back-off
of $\varpi<1$ reduces the effective DR by $-20\log_{10}(\varpi)$~dB
compared with the full-scale sinusoidal case.

When strong and weak components are simultaneously
present at the ADC input, the weaker component must be sufficiently
large to perturb the quantized representation of the stronger one.
Equivalently, its amplitude must induce at least a one-LSB change
in the quantizer output and only under this condition, both the components
can be resolved after quantization. Motivated by this requirement,
we derive a lower bound on the ADC resolution that guarantees resolvability
of the weaker component. To characterize the strongest and weakest
UL contributions that may coexist at the ADC input, we bound the received
DP and reflected powers. Let $\mathbf{H}_{m,k}^{\mathrm{DP}}=\alpha_{k}^{\mathrm{u}}\big(\widetilde{\widehat{\mathbf{H}}}_{m,k}+\widetilde{\widecheck{\mathbf{H}}}_{m,k}\big)$.
Using the inequality
\begin{align}
 & \mathbb{E}\Big\{\|\mathbf{H}_{m,k}^{\mathrm{DP}}\mathbf{x}_{m,k}^{(\ell)}\|^{2}\Big\}\nonumber \\
\leq\  & \|\mathbf{H}_{m,k}^{\mathrm{DP}}\|_{\mathrm{F}}^{2}\mathbb{E}\big\{\|\mathbf{x}_{m,k}^{(\ell)}\|^{2}\big\}=\|\mathbf{H}_{m,k}^{\mathrm{DP}}\|_{\mathrm{F}}^{2}\|\mathbf{F}_{m,k}\|_{\mathrm{F}}^{2},\label{eq:pow_ineq_1}
\end{align}
we obtain an upper bound on the received DP power.

\begin{figure*}[t]
\centering{}\subfloat[\label{fig:bit_bound_1}Quantized signal and quantization noise spectrum for $A_{\mathrm{s}}=0.9$ and $A_{\mathrm{w}}=0.05$.]{\begin{centering}
\includegraphics[width=0.28\textwidth,totalheight=3.3cm]{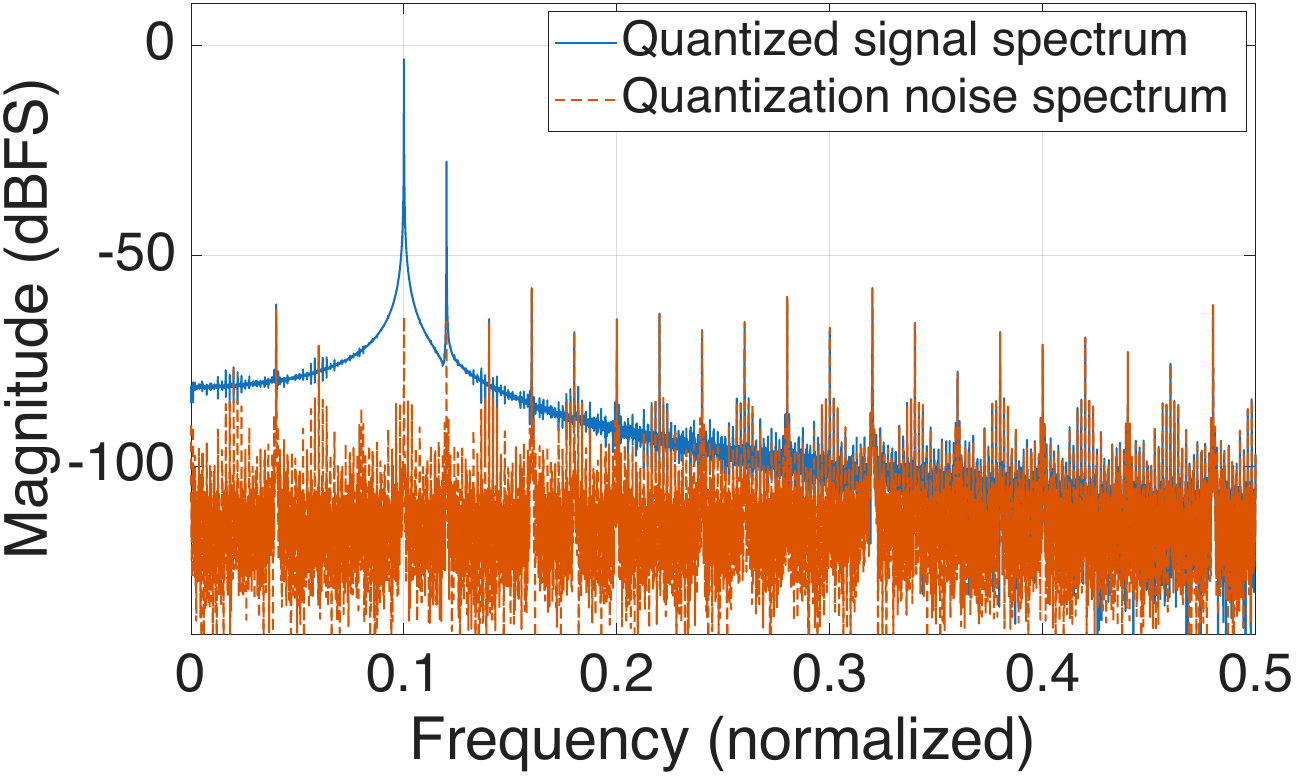}
\par\end{centering}
}\hfill{}\subfloat[\label{fig:bit_bound_2}Quantized signal and quantization noise spectrum
for $A_{\mathrm{s}}=0.9$ and $A_{\mathrm{w}}=10^{-4}$.]{\begin{centering}
\includegraphics[width=0.28\textwidth,totalheight=3.3cm]{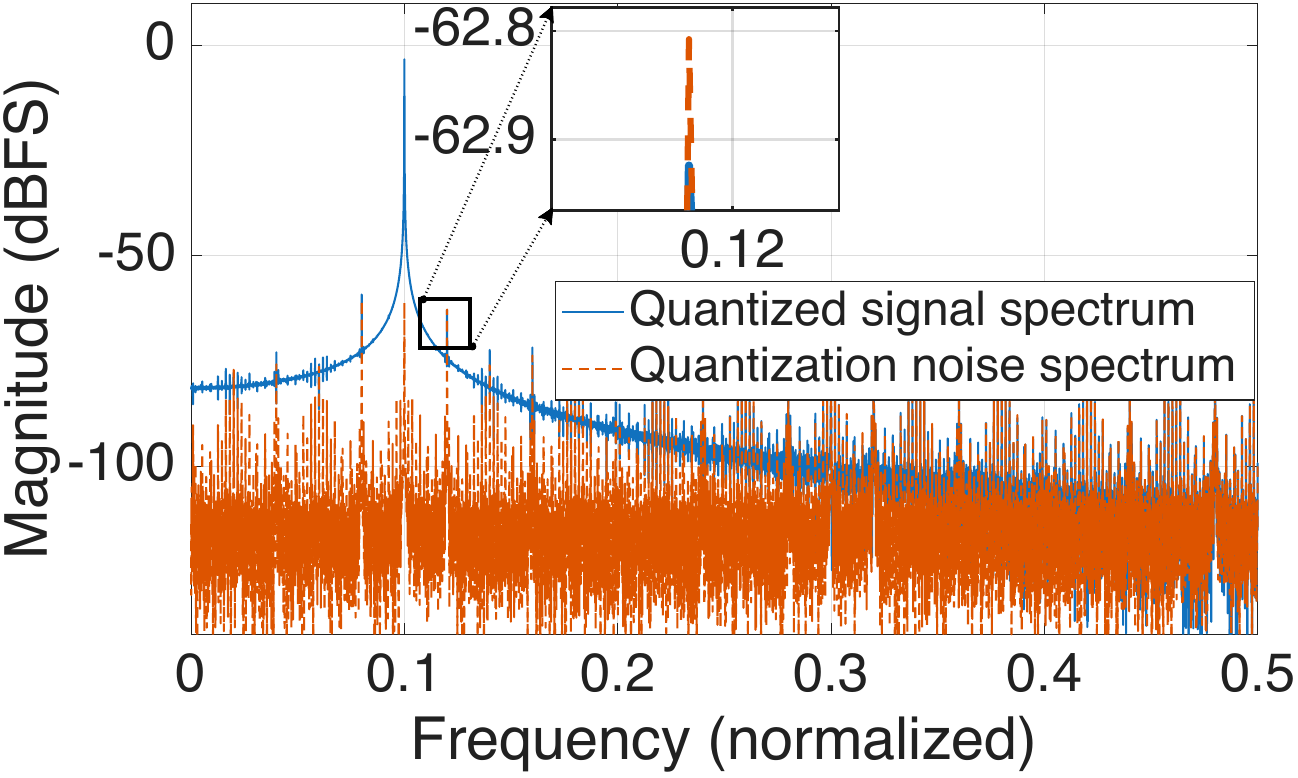}
\par\end{centering}
}\hfill{}\subfloat[\label{fig:bit_bound_3}Quantized signal and quantization noise spectrum
for $A_{\mathrm{s}}=0$ and $A_{\mathrm{w}}=10^{-4}$.]{\begin{centering}
\includegraphics[width=0.28\textwidth,totalheight=3.3cm]{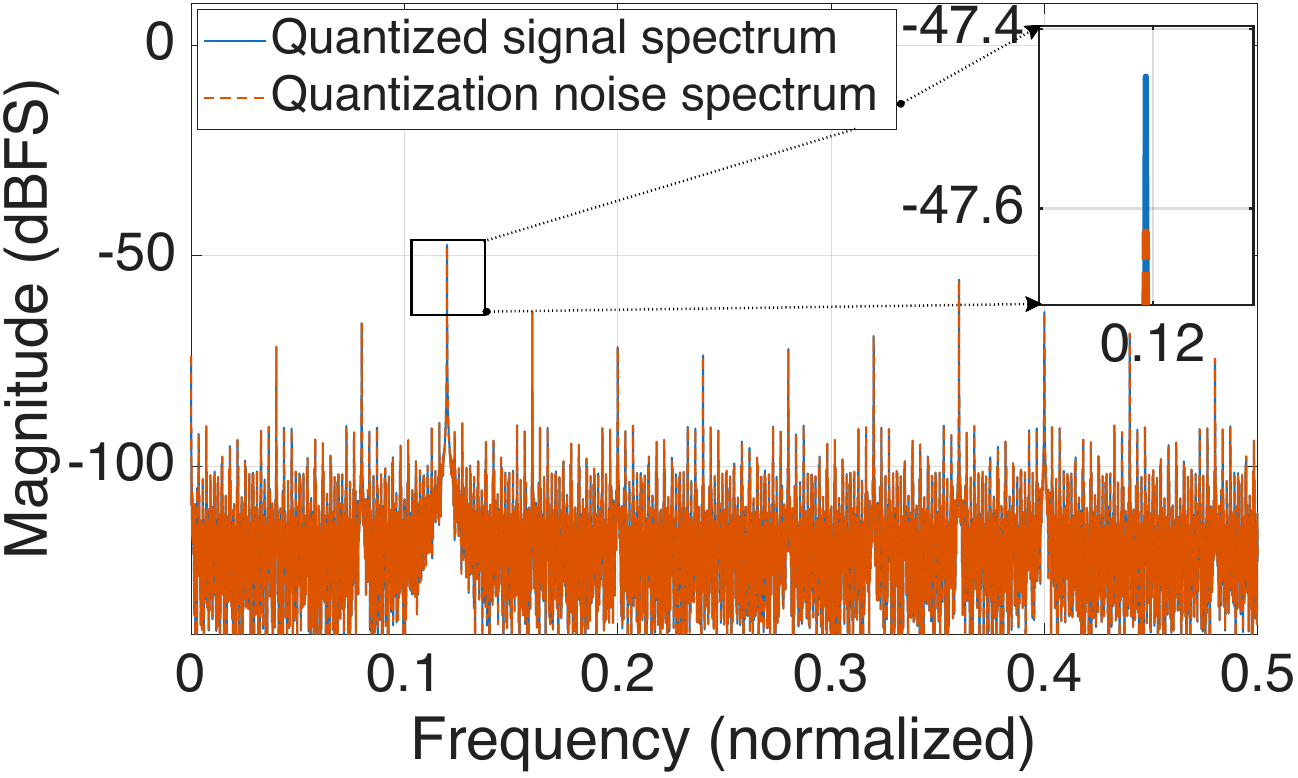}
\par\end{centering}
}\caption{\label{fig:bit_bound}Illustration of the dependence of the resolvability
of $x_{\mathrm{w}}$ in the presence of $x_{\mathrm{s}}$ on the ADC
resolution.}
\vspace{-0.2in}
\end{figure*}

The left hand side (LHS) of \eqref{eq:pow_ineq_1}
is the power of the signal received in the DP from the $k$-th UE
on subcarrier $m$ denoted as $P_{m,k}^{\mathrm{DP}}$. Let the total
received power in the DP from the $k$-th UE be $P_{k}^{\mathrm{DP}}=\sum_{m\in\mathcal{C}_{k}}P_{m,k}^{\mathrm{DP}}$.
From~\eqref{eq:pow_ineq_1}, $P_{k}^{\mathrm{DP}}\leq\sum_{m\in\mathcal{C}_{k}}\|\mathbf{H}_{m,k}^{\mathrm{DP}}\|_{\mathrm{F}}^{2}\|\mathbf{F}_{m,k}\|_{\mathrm{F}}^{2}$.
The DP channel in between the $k$-th UE and the BS is modeled as
a Rician channel. Therefore, $\|\mathbf{H}_{m,k}^{\mathrm{DP}}\|_{\mathrm{F}}^{2}$
can be computed as follows
\begin{align}
 & \|\mathbf{H}_{m,k}^{\mathrm{DP}}\|_{\mathrm{F}}^{2}=|\alpha_{k}^{\mathrm{u}}|^{2}\Big(\|\widetilde{\widehat{\mathbf{H}}}_{m,k}\|_{\mathrm{F}}^{2}+\|+\widetilde{\widecheck{\mathbf{H}}}_{m,k}\|_{\mathrm{F}}^{2}\nonumber \\
 & \quad\quad\qquad\qquad\qquad+2\Re\Big\{\langle\widetilde{\widehat{\mathbf{H}}}_{m,k},+\widetilde{\widecheck{\mathbf{H}}}_{m,k}\rangle_{\mathrm{F}}\Big\}\Big),\label{eq:H_dp_pow}\\
 & \|\widetilde{\widehat{\mathbf{H}}}_{m,k}\|_{\mathrm{F}}^{2}=\tfrac{\xi}{\xi+1}\|\mathbf{a}_{\mathrm{R}}(\theta_{k}^{\mathrm{r,u}})\mathbf{a}_{\mathrm{T},k}^{\mathrm{T}}(\theta_{k}^{\mathrm{t,u}})e^{-j2\pi m\Delta_{\mathrm{f}}\tau_{k}^{\mathrm{u}}}\|_{\mathrm{F}}^{2},\\
 & \|\widetilde{\widehat{\mathbf{H}}}_{m,k}\|_{\mathrm{F}}^{2}=\tfrac{\xi}{\xi+1},\|\widetilde{\widecheck{\mathbf{H}}}_{m,k}\|_{\mathrm{F}}^{2}=\tfrac{1}{\xi+1},\|\mathbf{H}_{m,k}^{\mathrm{DP}}\|_{\mathrm{F}}^{2}\approx|\alpha_{k}^{\mathrm{u}}|^{2}.\label{eq:los_pow}
\end{align}
The steering vectors are of unit-norm and therefore,~\eqref{eq:los_pow}
holds. Moreover, the standard Rician power split gives~\eqref{eq:los_pow}.
Moreover,~\eqref{eq:los_pow} is true because of the Rician modeling
of the DP channel. Under the Rician decomposition with unit-norm steering
vectors and with statistically independent/random phases across LoS
and NLoS components, the cross term $2\Re\big\{\langle\widetilde{\widehat{\mathbf{H}}}_{m,k},\widetilde{\widecheck{\mathbf{H}}}_{m,k}\rangle_{\mathrm{F}}\big\}\approx0$.
Hereafter, substituting $\|\mathbf{H}_{m,k}^{\mathrm{DP}}\|_{\mathrm{F}}^{2}$
into~\eqref{eq:pow_ineq_1} yields the compact bound
\begin{align}
P_{k}^{\mathrm{DP}}\leq\  & \sum\nolimits_{m\in\mathcal{C}_{k}}|\alpha_{k}^{\mathrm{u}}|^{2}\|\mathbf{F}_{m,k}\|_{\mathrm{F}}^{2}\nonumber \\
=\  & |\alpha_{k}^{\mathrm{u}}|^{2}\sum\nolimits_{m\in\mathcal{C}_{k}}\|\mathbf{F}_{m,k}\|_{\mathrm{F}}^{2}\leq|\alpha_{k}^{\mathrm{u}}|^{2}P_{k}^{\mathrm{u}}.\label{eq:dp_pow_3}
\end{align}
A parallel argument applies to the signal received through a reflection
involving $k$-th UE and $j$-th target, with effective reflected-channel
gain $\alpha_{k,j}^{\mathrm{u}}$. Denoting the corresponding received
reflected power by $P_{k,j}^{\mathrm{ref}}$, we obtain
\begin{equation}
P_{k,j}^{\mathrm{ref}}\leq|\alpha_{k,j}^{\mathrm{u}}|^{2}P_{k}^{\mathrm{u}}.\label{eq:ref_pow_1}
\end{equation}
The strongest UL component that may impinge on the BS receiver is
governed by the largest DP term, while the weakest component of interest
is typically the smallest target reflection. Using \eqref{eq:dp_pow_3}
and \eqref{eq:ref_pow_1}, we define
\begin{align}
P_{\mathrm{max}}^{\mathrm{DP}} & =\max_{k\in\mathcal{K}}|\alpha_{k}^{\mathrm{u}}|^{2}P_{k}^{\mathrm{u}},\label{eq:max_dp_pow}\\
P_{\mathrm{min}}^{\mathrm{ref}} & =\min_{k\in\mathcal{K}}\min_{j\in\mathcal{P}_{k}}|\alpha_{k,j}^{\mathrm{u}}|^{2}P_{k}^{\mathrm{u}}.\label{eq:min_ref_pow}
\end{align}
To avoid clipping, the ADC input limit must exceed the peak amplitude
of the strongest component after AGC scaling. A conservative amplitude
condition is, $A\geq\sqrt{P_{\mathrm{max}}^{\mathrm{DP}}}$. In reality,
the AGC circuit controls the input gain of the ADC through a variable
gain amplifier (VGA). This circuit adjusts the gain of the signal
such that the maximum and the minimum amplitudes of the signal fall
within the FSR of the ADC. To ensure that the weakest reflection is
not lost due to quantization, its amplitude at the ADC input should
be at least on the order of one quantization step. A conservative
resolvability condition is therefore
\begin{align}
 & \sqrt{P_{\mathrm{min}}^{\mathrm{ref}}}\geq\mathrm{LSB}\implies\sqrt{\min_{k\in\mathcal{K}}\min_{j\in\mathcal{P}_{k}}|\alpha_{k,j}^{\mathrm{u}}|^{2}P_{k}^{\mathrm{u}}}\geq\tfrac{2A}{2^{b}},\label{eq:P_ref_min}\\
 & b\geq\log_{2}\Big(2A/\sqrt{\min_{k\in\mathcal{K}}\min_{j\in\mathcal{P}_{k}}|\alpha_{k,j}^{\mathrm{u}}|^{2}P_{k}^{\mathrm{u}}}\Big).\label{eq:bit_bound}
\end{align}
From~\eqref{eq:bit_bound}, the central point is explicit that the
minimum required ADC resolution changes with the amplitude of the
weaker signal received at the BS. Consequently, if the reflected components
are extremely weak relative to the strongest DP, a finite-resolution
ADC may quantize the received signal in a way that removes the reflections,
thereby degrading HRF.

A simple illustration of the impact of ADC DR is
shown in Fig.~\ref{fig:bit_bound}. We consider a composite sinusoidal
signal $x(t)=x_{\mathrm{s}}(t)+x_{\mathrm{w}}(t)$, where $x_{\mathrm{s}}(t)=A_{\mathrm{s}}\sin(2\pi f_{1}t)$
and $x_{\mathrm{w}}(t)=A_{\mathrm{w}}\sin(2\pi f_{2}t)$ denote the
strong and weak components, respectively. An 8-bit uniform quantizer
with input range $[-A,A]$ and $A=1$ is used. The normalized frequencies
are set to $f_{1}=0.1F_{\mathrm{s}}$ and $f_{2}=0.12F_{\mathrm{s}}$,
where $F_{\mathrm{s}}$ denotes the sampling frequency. A $2^{16}$-point
FFT is applied to both the quantized signal and the corresponding
quantization noise. The resulting spectra are shown in Fig.~\ref{fig:bit_bound_1}--\ref{fig:bit_bound_3}.
In Fig.~\ref{fig:bit_bound_1}, for $A_{\mathrm{s}}=0.9$ and $A_{\mathrm{w}}=0.05$,
both tones are clearly visible above the quantization noise floor.
In contrast, Fig.~\ref{fig:bit_bound_2}, shows that when the weaker
amplitude is reduced to $A_{\mathrm{w}}=10^{-4}$, the corresponding
spectral component becomes buried under the quantization noise and
is no longer distinguishable. Fig.~\ref{fig:bit_bound_3} depicts
the spectrum of the weak tone in the absence of the strong component,
where it is again clearly observable. These results illustrate the
practical implication of~\eqref{eq:bit_bound} and reaffirm that
the resolvability of a weaker component is governed not by its absolute
power alone, but by its power relative to the simultaneously present
stronger component under a given ADC resolution.

\section{Optimization Framework for CRB-UL Rate Trade-off Characterization\label{sec:optimization_framework}}

In this section, we formulate two complementary optimization
problems to characterize the CRB–rate trade-off of HRF and to quantify
the impact of system parameters and ADC resolution. Unlike conventional
ISAC, where sensing and communication originate from a single transmitter,
HRF jointly exploits the BS probing signal and the UL signals of multiple
UEs, leading to trade-offs not captured by existing formulations.
We therefore consider a sensing-centric design~\eqref{eq:P0}, which
minimizes the CRB subject to per-UE rate constraints, and a communication-centric
design~\eqref{eq:P1}, which maximizes the UL sum rate subject to
a CRB constraint. Solving these problems traces the achievable CRB–rate
boundary of HRF under both sensing- and communication-driven operating
points while explicitly accounting for finite-resolution ADC effects
through the quantized CRB and rate expressions.

For tractability, we focus on the AoA vector $\boldsymbol{\theta}=[\theta_{1}^{\mathrm{tar}},\ldots,\theta_{P}^{\mathrm{tar}}]^{\mathrm{T}}$,
whose CRB follows from the FIM bound derived in Section~\ref{sec:FIM_lower_bound}.
The same framework applies to other parameters, but AoA estimation
suffices to expose the impact of ADC resolution on HRF. The sensing-
and communication-centric formulations and their solution methods
are detailed in Sections~\ref{subsec:Sensing-centric-design} and~\ref{subsec:Communication-centric-design},
respectively.

\subsection{Sensing-Centric Design \label{subsec:Sensing-centric-design}}

In this subsection, we develop a sensing-centric
design based on the optimization problem in \eqref{eq:P0}. Specifically,
we adopt the CRB expression in \eqref{eq:CRB_UB_LS} and specialize
it to AoA estimation by replacing the generic parameter vector $\bm{\psi}$
with $\bm{\theta}$. The resulting sensing-centric design problem
is formulated as
\begin{subequations}
\label{eq:P0}
\begin{align}
\!\!\!\underset{\mathbf{f}_{m},\{\mathbf{F}_{m,k}\}_{k\in\mathcal{K}}}{\minimize}\  & \trace\big(\widehat{\mathbf{C}}_{\bm{\theta}}\big(\mathbf{f}_{m},\mathbf{F}_{m,k}\big)\big)\label{eq:p0_obj_1}\\
\!\!\!\st\  & \mathrm{\trace}\Big(\sum\nolimits_{m\in\widehat{\mathcal{N}}}\mathbf{f}_{m}\mathbf{f}_{m}^{\mathrm{H}}\Big)\leq P_{\mathrm{BS}},\label{eq:p0_c_1}\\
 & \trace\Big(\sum\nolimits_{m\in\mathcal{C}_{k}}\mathbf{F}_{m,k}\mathbf{F}_{m,k}^{\mathrm{H}}\Big)\leq P_{k}^{\mathrm{u}}\ \forall k\in\mathcal{K},\label{eq:p0_c_2}\\
 & R_{k}\geq\Gamma_{k}\ \forall k\in\mathcal{K},\label{eq:p0_c_3}
\end{align}
\end{subequations}where $\widehat{\mathbf{C}}_{\bm{\theta}}\big(\mathbf{f}_{m},\mathbf{F}_{m,k}\big)=\big(\widehat{\mathbf{G}}_{\bm{\theta}}(\mathbf{f}_{m},\mathbf{F}_{m,k})\big)^{-1}$,
$\Gamma_{k}$ is the rate threshold for $k^{\mathrm{th}}$ UE. The
$(i,j)^{\mathrm{th}}$ element of $\widehat{\mathbf{G}}_{\bm{\theta}}(\mathbf{f}_{m},\mathbf{F}_{m,k})$
can be computed from \eqref{eq:FIM_LB_2} by replacing $\bm{\psi}$
with $\bm{\theta}$ and $\psi_{i}$ and $\psi_{j}$ with $\theta_{i}^{\mathrm{tar}}$
and $\theta_{j}^{\mathrm{tar}}$, respectively. Furthermore, $\frac{\partial\tilde{\mathbf{x}}_{m}^{(\ell)}}{\partial\theta_{i}^{\mathrm{tar}}}$
is needed to compute $\widehat{\mathbf{G}}_{\bm{\theta}}(\mathbf{f}_{m},\mathbf{F}_{m,k})$,
which can be expressed in closed form as
\[
\frac{\partial\tilde{\mathbf{x}}_{m}^{(\ell)}}{\partial\theta_{i}^{\mathrm{tar}}}=\begin{cases}
\alpha_{i}\phi_{i}(\ell,m)\mathbf{A}_{i}^{\mathrm{tar}}\mathbf{f}_{m}b_{m,0}^{(\ell)}, & \mathrm{for}\,m\in\widecheck{\mathcal{N}}\\
\alpha_{k,i}^{\mathrm{u}}\phi_{k,i}(\ell,m)\mathbf{A}_{k,i}^{\mathrm{tar}}\mathbf{F}_{m,k}\mathbf{b}_{m,k}^{(\ell)}, & \mathrm{for}\,m\in\widehat{\mathcal{N}},\,i\in\mathcal{P}_{k}
\end{cases}
\]
where $\phi_{i}(\ell,m)=e^{-j2\pi(m\Delta_{\mathrm{f}}\tau_{i}^{\mathrm{tar}}-f_{\mathrm{d},i}^{\mathrm{tar}}\ell T_{\mathrm{tot}})}$,
$\mathbf{A}_{i}^{\mathrm{tar}}=\mathbf{a}_{\mathrm{R}}'(\theta_{i}^{\mathrm{tar}})\mathbf{a}_{\mathrm{T}}^{\mathrm{T}}(\theta_{i}^{\mathrm{tar}})+\mathbf{a}_{\mathrm{R}}(\theta_{i}^{\mathrm{tar}})\big(\mathbf{a}_{\mathrm{T}}'(\theta_{i}^{\mathrm{tar}})\big)^{\mathrm{T}}$,
$\phi_{k,i}(\ell,m)=e^{-j2\pi(m\Delta_{\mathrm{f}}\tau_{k,i}^{\mathrm{u}}-f_{\mathrm{d},i}^{\mathrm{tar}}\ell T_{\mathrm{tot}})}$,
and $\mathbf{A}_{k,i}^{\mathrm{tar}}=\mathbf{a}_{\mathrm{R}}'(\theta_{i}^{\mathrm{tar}})\mathbf{a}_{\mathrm{T},k}^{\mathrm{T}}(\theta_{k,i}^{\mathrm{u}})$.
The objective in \eqref{eq:p0_obj_1} and the power constraints in
\eqref{eq:p0_c_1} and \eqref{eq:p0_c_2} are quadratic in $\mathbf{f}_{m}$
and $\mathbf{F}_{m,k}$. To obtain an SDP form suitable for CVX, we
rewrite the FIM terms so that the optimization variables appear through
outer products $\mathbf{f}_{m}\mathbf{f}_{m}^{\mathrm{H}}$ and $\mathbf{f}_{m,k}\mathbf{f}_{m,k}^{\mathrm{H}}$,
where $\mathbf{f}_{m,k}\triangleq\mathrm{vec}(\mathbf{F}_{m,k})$.

For $m\in\widecheck{\mathcal{N}}$, the product $\bigg(\tfrac{\partial\tilde{\mathbf{x}}_{m}^{(\ell)}}{\partial\theta_{i}^{\mathrm{tar}}}\bigg)^{\mathrm{H}}\widehat{\mathbf{R}}_{\tilde{\mathbf{e}}^{(\ell)}\tilde{\mathbf{e}}^{(\ell)}}^{-1}[m]\Big(\tfrac{\partial\tilde{\mathbf{x}}_{m}^{(\ell)}}{\partial\theta_{j}^{\mathrm{tar}}}\Big)$
appearing in the FIM can be recast using the trace operator as follows
\begin{align}
 & \Big(\tfrac{\partial\tilde{\mathbf{x}}_{m}^{(\ell)}}{\partial\theta_{i}^{\mathrm{tar}}}\Big)^{\mathrm{H}}\widehat{\mathbf{R}}_{\tilde{\mathbf{e}}^{(\ell)}\tilde{\mathbf{e}}^{(\ell)}}^{-1}[m]\tfrac{\partial\tilde{\mathbf{x}}_{m}^{(\ell)}}{\partial\theta_{j}^{\mathrm{tar}}}\nonumber \\
=\  & \trace\Big[(\alpha_{i}\phi_{i}(\ell,m)\mathbf{A}_{i}^{\mathrm{tar}}\mathbf{f}_{m}b_{m,0}^{(\ell)})^{\mathrm{H}}\nonumber \\
 & \times\widehat{\mathbf{R}}_{\tilde{\mathbf{e}}^{(\ell)}\tilde{\mathbf{e}}^{(\ell)}}^{-1}[m]\alpha_{j}\phi_{j}(\ell,m)\mathbf{A}_{j}^{\mathrm{tar}}\mathbf{f}_{m}b_{m,0}^{(\ell)}\Big]\nonumber \\
=\  & \trace\Big(\beta_{i,j}(\ell,m)\mathbf{R}_{m}(\mathbf{A}_{i}^{\mathrm{tar}})^{\mathrm{H}}\widehat{\mathbf{R}}_{\tilde{\mathbf{e}}^{(\ell)}\tilde{\mathbf{e}}^{(\ell)}}^{-1}[m]\mathbf{A}_{j}^{\mathrm{tar}}\Big),\label{eq:DL_FIM_trace_4}
\end{align}
where $\beta_{i,j}(\ell,m)=\alpha_{i}^{*}\alpha_{j}\phi_{i}^{*}(\ell,m)\phi_{j}(\ell,m)|b_{m,0}^{(\ell)}|^{2}$,
$\mathbf{R}_{m}=\mathbf{f}_{m}\mathbf{f}_{m}^{\mathrm{H}}$. This
follows from the facts that the quadratic form is scalar (hence equal
to its trace) and that the trace is invariant under cyclic permutations.

For $m\in\widehat{\mathcal{N}}$, an analogous manipulation
is applied, which is written as follows
\begin{align}
 & \Big(\tfrac{\partial\tilde{\mathbf{x}}_{m}^{(\ell)}}{\partial\theta_{i}^{\mathrm{tar}}}\Big)^{\mathrm{H}}\widehat{\mathbf{R}}_{\tilde{\mathbf{e}}^{(\ell)}\tilde{\mathbf{e}}^{(\ell)}}^{-1}[m]\tfrac{\partial\tilde{\mathbf{x}}_{m}^{(\ell)}}{\partial\theta_{j}^{\mathrm{tar}}}=\trace\Big[\beta_{k,i,j}^{\mathrm{u}}(\ell,m)\nonumber \\
 & \times(\mathbf{A}_{k,i}^{\mathrm{tar}}\mathbf{F}_{m,k}\mathbf{b}_{m,k}^{(\ell)})^{\mathrm{H}}\widehat{\mathbf{R}}_{\tilde{\mathbf{e}}^{(\ell)}\tilde{\mathbf{e}}^{(\ell)}}^{-1}[m]\mathbf{A}_{k,j}^{\mathrm{tar}}\mathbf{F}_{m,k}\mathbf{b}_{m,k}^{(\ell)}\Big],\label{eq:UL_FIM_trace_1}
\end{align}
where $\beta_{k,i,j}^{\mathrm{u}}(\ell,m)=\alpha_{k,i}^{\mathrm{u}*}\alpha_{k,j}^{\mathrm{u}}\phi_{k,i}^{*}(\ell,m)\phi_{k,j}(\ell,m)$.
The term $\mathbf{F}_{m,k}\mathbf{b}_{m,k}^{(\ell)}$ in \eqref{eq:UL_FIM_trace_1}
can be written as follows $\mathbf{B}_{m,k}^{(\ell)}\mathbf{f}_{m,k}$,
where $\mathbf{B}_{m,k}^{(\ell)}=\big(\mathbf{b}_{m,k}^{(\ell)}\big)^{\mathrm{T}}\otimes\mathbf{I}_{N_{k}^{\mathrm{u}}}$
and $\mathbf{f}_{m,k}=\mathrm{vec}(\mathbf{F}_{m,k})$. Substitute
the term $\mathbf{F}_{m,k}\mathbf{b}_{m,k}^{(\ell)}$ with $\mathbf{B}_{m,k}^{(\ell)}\mathbf{f}_{m,k}$
in \eqref{eq:UL_FIM_trace_1}, we get
\begin{align}
 & \Big(\tfrac{\partial\tilde{\mathbf{x}}_{m}^{(\ell)}}{\partial\theta_{i}^{\mathrm{tar}}}\Big)^{\mathrm{H}}\widehat{\mathbf{R}}_{\tilde{\mathbf{e}}^{(\ell)}\tilde{\mathbf{e}}^{(\ell)}}^{-1}[m]\tfrac{\partial\tilde{\mathbf{x}}_{m}^{(\ell)}}{\partial\theta_{j}^{\mathrm{tar}}}=\trace\Big[\beta_{k,i,j}^{\mathrm{u}}(\ell,m)\mathbf{R}_{m,k}\nonumber \\
 & \qquad\quad\times(\mathbf{B}_{m,k}^{(\ell)})^{\mathrm{H}}(\mathbf{A}_{k,i}^{\mathrm{tar}})^{\mathrm{H}}\widehat{\mathbf{R}}_{\tilde{\mathbf{e}}^{(\ell)}\tilde{\mathbf{e}}^{(\ell)}}^{-1}[m]\mathbf{A}_{k,j}^{\mathrm{tar}}\mathbf{B}_{m,k}^{(\ell)}\Big],\label{eq:UL_FIM_trace_2}
\end{align}
where $\mathbf{R}_{m,k}=\mathbf{f}_{m,k}\mathbf{f}_{m,k}^{\mathrm{H}}$.
Using these rewritten DL and UL contributions, the modified $(i,j)^{\mathrm{th}}$
elements of the FIM and CRB are expressed as
\begin{align}
 & [\mathbf{\tilde{G}}_{\bm{\theta}}]_{i,j}=\sum\nolimits_{\ell=0}^{N_{\mathrm{s}}-1}\Big[2\eta^{2}\Re\Big\{\sum\nolimits_{m\in\widecheck{\mathcal{N}}}\Big(\tfrac{\partial\tilde{\mathbf{x}}_{m}^{(\ell)}}{\partial\theta_{i}^{\mathrm{tar}}}\Big)^{\mathrm{H}}\widehat{\mathbf{R}}_{\tilde{\mathbf{e}}^{(\ell)}\tilde{\mathbf{e}}^{(\ell)}}^{-1}[m]\nonumber \\
 & \times\tfrac{\partial\tilde{\mathbf{x}}_{m}^{(\ell)}}{\partial\theta_{j}^{\mathrm{tar}}}+\sum\nolimits_{m\in\widehat{\mathcal{N}}}\Big(\tfrac{\partial\tilde{\mathbf{x}}_{m}^{(\ell)}}{\partial\theta_{i}^{\mathrm{tar}}}\Big)^{\mathrm{H}}\widehat{\mathbf{R}}_{\tilde{\mathbf{e}}^{(\ell)}\tilde{\mathbf{e}}^{(\ell)}}^{-1}[m]\tfrac{\partial\tilde{\mathbf{x}}_{m}^{(\ell)}}{\partial\theta_{j}^{\mathrm{tar}}}\Big\}\Big],\label{eq:mod_FIM_LB_2}
\end{align}
\begin{equation}
[\tilde{\mathbf{C}}_{\bm{\theta}}]_{i,j}=[(\mathbf{\tilde{G}}_{\bm{\theta}})^{-1}]_{i,j}.\label{eq:mod_CRB_UB_LS}
\end{equation}
The expression of $\Big(\tfrac{\partial\tilde{\mathbf{x}}_{m}^{(\ell)}}{\partial\theta_{i}^{\mathrm{tar}}}\Big)^{\mathrm{H}}\widehat{\mathbf{R}}_{\tilde{\mathbf{e}}^{(\ell)}\tilde{\mathbf{e}}^{(\ell)}}^{-1}[m]\tfrac{\partial\tilde{\mathbf{x}}_{m}^{(\ell)}}{\partial\theta_{j}^{\mathrm{tar}}}$
for $m\in\widecheck{\mathcal{N}}$ and for $m\in\widehat{\mathcal{N}}$
is given in \eqref{eq:DL_FIM_trace_4} and \eqref{eq:UL_FIM_trace_2},
respectively.

We have to note that the rate expression given in
\eqref{eq:rate} contains the term $\mathbf{F}_{m,k}\mathbf{F}_{m,k}^{\mathrm{H}}$.
This has to be written in terms of $\mathbf{R}_{m,k}$. For each user
$k$ and $\forall\;r\in\{1,\ldots,S_{k}\}$, we define a basis vector
$\mathbf{e}_{r}^{(k)}\in\mathbb{R}^{S_{k}}$ as follows
\[
[\mathbf{e}_{r}^{(k)}]_{i}=\begin{cases}
1, & i=r\\
0. & i\neq r
\end{cases}
\]
Using $\mathbf{e}_{r}^{(k)}$, we define a column selection matrix
as $\mathbf{E}_{r}^{(k)}\triangleq(\mathbf{e}_{r}^{(k)})^{\mathrm{T}}\otimes\mathbf{I}_{N_{k}^{\mathrm{u}}}$
so that $\mathbf{E}_{r}^{(k)}\mathbf{f}_{m,k}=\mathbf{F}_{m,k}(:,r)=\mathbf{f}_{m,k}^{(r)}$.
Using this, $\mathbf{F}_{m,k}\mathbf{F}_{m,k}^{\mathrm{H}}$ can be
written as follows
\begin{align}
 & \mathbf{F}_{m,k}\mathbf{F}_{m,k}^{\mathrm{H}}\nonumber \\
=\  & \sum\nolimits_{r=1}^{S_{k}}\mathbf{E}_{r}^{(k)}\mathbf{f}_{m,k}(\mathbf{E}_{r}^{(k)}\mathbf{f}_{m,k})^{\mathrm{H}}=\sum\nolimits_{r=1}^{S_{k}}\mathbf{E}_{r}^{(k)}\mathbf{R}_{m,k}\mathbf{E}_{r}^{(k)\mathrm{H}}.\label{eq:mod_user_precoder_cov}
\end{align}
Substituting \eqref{eq:mod_user_precoder_cov} into the rate expression
in \eqref{eq:rate} yields a reformulated rate constraint of the form
\begin{align}
 & \tilde{R}_{k}=\sum\nolimits_{m\in\mathcal{C}_{k}}\ln\,\mathrm{det}\Big[\mathbf{I}_{N_{\mathrm{r}}^{\mathrm{bs}}}+\eta^{2}\Tilde{\mathbf{H}}_{\ell,m,k}^{\mathrm{UL}}\nonumber \\
 & \times\Big(\sum\nolimits_{r=1}^{S_{k}}\mathbf{E}_{r}^{(k)}\mathbf{R}_{m,k}\mathbf{E}_{r}^{(k)\mathrm{H}}\Big)\big(\Tilde{\mathbf{H}}_{\ell,m,k}^{\mathrm{UL}}\big)^{\mathrm{H}}\widehat{\mathbf{R}}_{\tilde{\mathbf{e}}^{(\ell)}\tilde{\mathbf{e}}^{(\ell)}}^{-1}[m]\Big].\label{eq:mod_rate}
\end{align}

Similarly, the UL power constraint in \eqref{eq:p0_c_2}
can be written directly in terms of $\mathbf{R}_{m,k}$. Since $\mathbf{f}_{m,k}=\mathrm{vec}(\mathbf{F}_{m,k})$,
we have $\trace(\mathbf{f}_{m,k}^{\mathrm{H}}\mathbf{f}_{m,k})=\sum_{i}\sum_{j}|[\mathbf{F}_{m,k}]_{i,j}|^{2}=\|\mathbf{F}_{m,k}\|_{\mathrm{F}}^{2}=\trace(\mathbf{F}_{m,k}\mathbf{F}_{m,k}^{\mathrm{H}})$.
Therefore, $\trace\Big(\!\sum_{m\in\mathcal{C}_{k}}\!\!\mathbf{F}_{m,k}\mathbf{F}_{m,k}^{\mathrm{H}}\Big)=\trace\Big(\!\sum_{m\in\mathcal{C}_{k}}\!\!\mathbf{f}_{m,k}\mathbf{f}_{m,k}^{\mathrm{H}}\Big)=\trace\Big(\!\sum_{m\in\mathcal{C}_{k}}\!\!\mathbf{R}_{m,k}\Big)$.
We handle the term $\trace(\widehat{\mathbf{C}}_{\bm{\theta}})=\trace\!\big(\tilde{\mathbf{G}}_{\boldsymbol{\theta}}^{-1}\big)$
by introducing an epigraph variable. Let $\mathbf{W}\in\mathbb{C}^{P\times P}$
be a Hermitian matrix and impose the inequality $\mathbf{W}\succeq\tilde{\mathbf{G}}_{\boldsymbol{\theta}}^{-1}$.
Then $\trace(\mathbf{W})$ is an upper bound on $\trace\!\big(\tilde{\mathbf{G}}_{\boldsymbol{\theta}}^{-1}\big)$.
Minimizing $\trace(\mathbf{W})$ therefore minimizes this upper bound.
At the optimum, the inequality becomes tight because any strictly
larger $\mathbf{W}$ would increase the trace without improving feasibility.
This yields $\trace(\mathbf{W}^{\star})=\trace\!\big(\tilde{\mathbf{G}}_{\boldsymbol{\theta}}^{-1}\big)$.
The constraint $\mathbf{W}\succeq\tilde{\mathbf{G}}_{\boldsymbol{\theta}}^{-1}$
can be written as a linear matrix inequality (LMI) using the Schur
complement. Since $\tilde{\mathbf{G}}_{\boldsymbol{\theta}}\succ\mathbf{0}$,
the following equivalence holds
\begin{equation}
\begin{bmatrix}\mathbf{W} & \mathbf{I}_{P}\\
\mathbf{I}_{P} & \tilde{\mathbf{G}}_{\boldsymbol{\theta}}
\end{bmatrix}\succeq\mathbf{0}\Longleftrightarrow\mathbf{W}-\tilde{\mathbf{G}}_{\boldsymbol{\theta}}^{-1}\succeq\mathbf{0}\Longleftrightarrow\mathbf{W}\succeq\tilde{\mathbf{G}}_{\boldsymbol{\theta}}^{-1}.
\end{equation}
This converts the matrix inverse inside the objective into an LMI
constraint. As a result, minimizing $\trace(\mathbf{W})$ subject
to this LMI is equivalent to minimizing $\trace\!\big(\tilde{\mathbf{G}}_{\boldsymbol{\theta}}^{-1}\big)$,
while keeping the formulation convex.

Collecting these transformations, \eqref{eq:P0}
can be equivalently rewritten in terms of variables $\mathbf{R}_{m}$
and $\mathbf{R}_{m,k}$ as {\color{blue}}
\begin{subequations}
\label{eq:P0_1}
\begin{align}
\underset{\mathbf{R}_{m},\{\mathbf{R}_{m,k}\}_{k\in\mathcal{K}}}{\minimize}\  & \trace\big(\mathbf{W}\big)\label{eq:p0_1_obj_1}\\
\st\  & \begin{bmatrix}\mathbf{W} & \mathbf{I}_{P}\\
\mathbf{I}_{P} & \tilde{\mathbf{G}}_{\bm{\theta}}(\mathbf{R}_{m},\mathbf{R}_{m,k})
\end{bmatrix}\succcurlyeq0,\label{eq:p0_1_c_1}\\
 & \mathrm{\trace}\Big(\sum\nolimits_{m\in\widecheck{\mathcal{N}}}\mathbf{R}_{m}\Big)\leq P_{\mathrm{BS}},\label{eq:p0_1_c_2}\\
 & \trace\Big(\sum\nolimits_{m\in\mathcal{C}_{k}}\mathbf{R}_{m,k}\Big)\leq P_{k}^{\mathrm{u}}\ \forall k\in\mathcal{K},\label{eq:p0_1_c_3}\\
 & \tilde{R}_{k}(\mathbf{R}_{m,k})\geq\Gamma_{k}\quad\forall k\in\mathcal{K},\label{eq:p0_1_c_4}\\
 & \operatorname{rank}(\mathbf{R}_{m})=1\ \forall m,\label{eq:p0_1_c_5}\\
 & \operatorname{rank}(\mathbf{R}_{m,k})=S_{k}\ \forall m,k\label{eq:p0_1_c_6}
\end{align}
\end{subequations}
The reformulation introduces rank constraints,
\eqref{eq:p0_1_c_5}, and \eqref{eq:p0_1_c_6}, which preserve equivalence
to the original precoder variables. These rank conditions render the
problem non-convex. Following a standard semidefinite relaxation (SDR)
approach, we drop the rank constraints and obtain a convex semidefinite
program (SDP). Thus, the problem in \eqref{eq:P0_1} can be reformulated
as follows
\begin{subequations}
\label{eq:P0_2}
\begin{align}
\underset{\mathbf{R}_{m},\{\mathbf{R}_{m,k}\}_{k\in\mathcal{K}}}{\minimize}\  & \trace\big(\mathbf{W}\big)\label{eq:p0_2_obj_1}\\
\st\  & \eqref{eq:p0_1_c_1},\eqref{eq:p0_1_c_2},\eqref{eq:p0_1_c_3},\eqref{eq:p0_1_c_4}.\nonumber 
\end{align}
\end{subequations}
The resulting problem in \eqref{eq:P0_2} is convex
and therefore directly solvable using CVX or any other convex optimization
package. To recover $\mathbf{f}_{m}$ from $\mathbf{R}_{m}$ ($\forall m$),
we apply eigenvalue decomposition when the rank-one condition in~\eqref{eq:p0_1_c_5}
is satisfied. Otherwise, we employ Gaussian randomization to generate
feasible rank-one precoding vectors~\cite[Sec. V]{Rank_1}. A similar
routine is followed to recover $\mathbf{F}_{m,k}$ ($\forall m,k$)
from $\mathbf{R}_{m,k}$.

\subsection{Communication-Centric Design\label{subsec:Communication-centric-design}}

In this section, we consider a communication-centric
design problem as given in \eqref{eq:P1}. The goal is to maximize
the aggregate UL rate across all UEs while ensuring a prescribed sensing
quality, enforced through a CRB constraint. Using the rate expression
$\tilde{R}_{k}(\mathbf{R}_{m,k})$ in \eqref{eq:mod_rate}, the communication-centric problem can be written as
\begin{subequations}
\label{eq:P1}
\begin{flalign}
\underset{\mathbf{R}_{m},\{\mathbf{R}_{m,k}\}_{k\in\mathcal{K}}}{\maximize}\; & \sum\nolimits_{k\in\mathcal{K}}\tilde{R}_{k}(\mathbf{R}_{m,k})\label{eq:p1_obj_1}\\
\st\; & \eqref{eq:p0_1_c_1},\eqref{eq:p0_1_c_2},\eqref{eq:p0_1_c_3},\eqref{eq:p0_1_c_5},\eqref{eq:p0_1_c_6},\nonumber \\
\; & \trace\big(\mathbf{W}\big)\leq\mu,\label{eq:p1_c3}
\end{flalign}
\end{subequations}
where $\mu$ is the threshold for the trace of the
CRB matrix. The problem formulated in \eqref{eq:P1} is non-convex
due to~\eqref{eq:p0_1_c_5} and~\eqref{eq:p0_1_c_6}. Following the same SDR strategy adopted in the sensing-centric design, we drop
these rank constraints to obtain the relaxed formulation
\begin{subequations}
\label{eq:P1_1}
\begin{flalign}
\underset{\mathbf{R}_{m},\{\mathbf{R}_{m,k}\}_{k\in\mathcal{K}}}{\maximize}\; & \sum\nolimits_{k\in\mathcal{K}}\tilde{R}_{k}(\mathbf{R}_{m,k})\label{eq:p1_1_obj_1}\\
\st\; & \eqref{eq:p0_1_c_1},\eqref{eq:p0_1_c_2},\eqref{eq:p0_1_c_3},\nonumber \\
\; & \trace\big(\mathbf{W}\big)\leq\mu.\label{eq:p1_1_c_2}
\end{flalign}
\end{subequations}
The problem in \eqref{eq:P1_1} is convex and can
be solved using CVX. In the next section, we solve the sensing- and
communication-centric optimization problems to generate numerical
results and to draw insights into the impact of ADC resolution on
HRF performance. In particular, the two formulations allow us to trace
CRB–rate trade-off curves and highlight how finite-resolution effects
influence the achievable sensing–communication operating points.

\subsection{Computational Complexity}

From~\eqref{eq:P0_2}, we observe that, in the real
domain, the total number of optimization variables is given by $P^{2}+N_{\mathrm{c}}(N_{\mathrm{t}}^{\mathrm{bs}})^{2}+\sum_{k\in\mathcal{K}}|\mathcal{C}_{k}|(N_{k}^{\mathrm{u}}S_{k})^{2}+N_{\mathrm{c}}((2N_{\mathrm{r}}^{\mathrm{bs}})^{2}+3N_{\mathrm{r}}^{\mathrm{bs}}+1)$
while the total number of SDP constraints is $1+2N_{\mathrm{c}}$.
Hence, by using the standard interior-point complexity scaling in~\cite[Sec. 6.6.3]{ben-tal_lectures_2001},
the overall arithmetic complexity for solving~\eqref{eq:P0_2} is
given by
\begin{align}
\mathcal{C}_{\mathrm{compl}} & \approx\mathcal{O}\Big([P^{2}\!+\!N_{\mathrm{c}}(N_{\mathrm{t}}^{\mathrm{bs}})^{2}\!+\!\sum\nolimits_{k\in\mathcal{K}}\!|\mathcal{C}_{k}|(N_{k}^{\mathrm{u}}S_{k})^{2}]\nonumber \\
 & \times(N_{\mathrm{r}}^{\mathrm{bs}})^{3.5}\!N_{\mathrm{c}}^{1.5}\!+\!(N_{\mathrm{r}}^{\mathrm{bs}})^{5.5}\!N_{\mathrm{c}}^{2.5}\Big).\label{eq:comp_complexity}
\end{align}
By following the same steps, it can be shown that the problem in~\eqref{eq:P1_1}
admits the same order of complexity as in~\eqref{eq:comp_complexity}.

\section{Numerical Results and Discussion\label{sec:results}}

In this section, we quantify how the ADC resolution
prescribed by~\eqref{eq:bit_bound}, together with key system parameters,
shapes the sensing–communication trade-off of the HRF architecture.
Sensing performance is measured via the CRB and communication performance
via the achievable spectral efficiency (SE), obtained by solving~\eqref{eq:P0_2}
and~\eqref{eq:P1_1}, respectively.

Unless otherwise stated, the simulations consider
a BS with $N_{\mathrm{t}}^{\mathrm{bs}}=N_{\mathrm{r}}^{\mathrm{bs}}=8$
antennas, $K=2$ UL UEs each equipped with $N_{k}^{\mathrm{u}}=4$
transmit antennas, and $P=2$ targets. The carrier frequency is $f_{\mathrm{c}}=27$~GHz,
with $N_{\mathrm{c}}=64$ subcarriers and $N_{\mathrm{s}}=14$ OFDM
symbols in both UL and DL. Each UE occupies $N_{\mathrm{c}}/K$ subcarriers
with spacing $240$~kHz. The BS transmit power is $10$~dBm, the
noise PSD is $-174$~dBm/Hz, and the receiver noise figure is $9$~dB.
All targets have radar cross section $0.1\,\mathrm{m}^{2}$. All links,
including the Rician-faded UL DPs, follow the wideband channel models
in~\eqref{eq:fd_echo_chan} and~\eqref{eq:fd_ul_chan}, with Rician
factor $\xi=3$~dB and path-loss computed as in~\cite{HRF}. The
data symbols $b_{m,0}^{(\ell)}$ and $\mathbf{b}_{m,k}^{(\ell)}$
are drawn from unit-energy 16-QAM constellations, and each UE transmits
$\min(N_{\mathrm{r}}^{\mathrm{bs}},N_{k}^{\mathrm{u}})$ streams.
The BS is located at $(0,0,10)$, while targets and users are uniformly
distributed over disks of radius $10$~m centered at $(10,10,10)$
and $(20,20,0)$, respectively. For each configuration,~\eqref{eq:P0_2}
and~\eqref{eq:P1_1} are solved over $100$ independent realizations
of UE and target locations, and the reported CRB and SE values are
obtained by averaging across these realizations, capturing the typical
HRF behavior.
\begin{figure*}[tbh]
\begin{minipage}[t]{0.32\textwidth}%
\begin{center}
\includegraphics[width=1\columnwidth,totalheight=4.5cm]{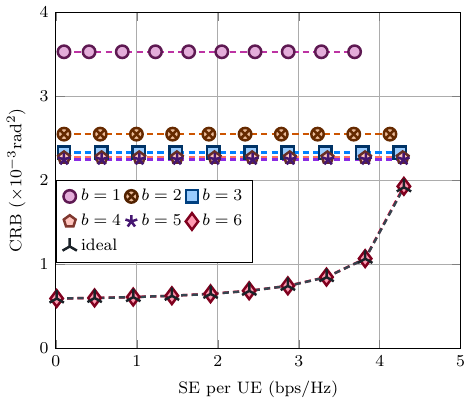}\vspace{-0.2in}
\par\end{center}
\caption{The impact of the ADC DR given in~\eqref{eq:bit_bound} on the HRF
system from the perspective of CRB-rate Pareto boundary obtained through~\eqref{eq:P0_2}.}
\label{fig:results_bit_bound_1}%
\end{minipage}\hfill{}%
\begin{minipage}[t]{0.32\textwidth}%
\begin{center}
\includegraphics[width=1\columnwidth,totalheight=4.5cm]{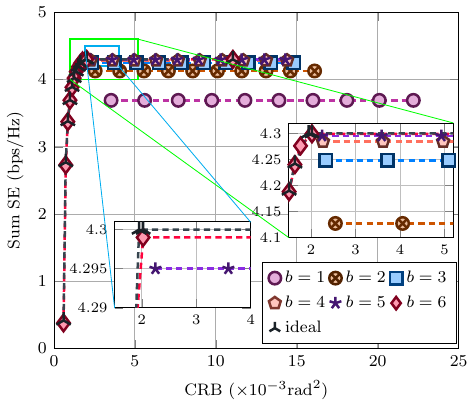}\vspace{-0.2in}
\par\end{center}
\caption{The impact of the ADC DR given in~\eqref{eq:bit_bound} on the HRF
system from the perspective of CRB-rate Pareto boundary obtained through~\eqref{eq:P1_1}.}
\label{fig:results_bit_bound_2}%
\end{minipage}\hfill{}%
\begin{minipage}[t]{0.32\textwidth}%
\begin{center}
\includegraphics[width=1\columnwidth,totalheight=5cm,keepaspectratio]{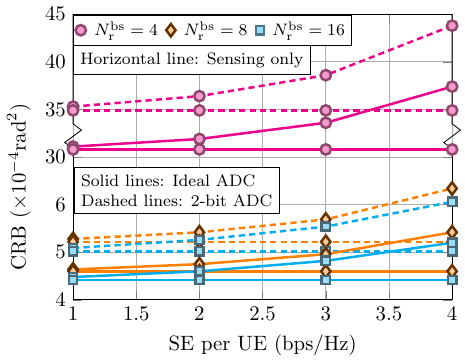}\vspace{-0.2in}
\par\end{center}
\caption{CRB-rate trade-off computed using optimization problem in \eqref{eq:P0_2}
for three different receive antenna configurations.}
\label{fig:crb_vs_rate_1}%
\end{minipage}
\vspace{-0.1in}
\end{figure*}

\subsection{Impact of ADC DR on the CRB-rate Pareto Boundary}

The minimum ADC resolution required to resolve weak
target reflections in the presence of strong UE DPs was established
in~\eqref{eq:fd_ul_chan}. When this condition is violated, the reflected
components fall below the quantization noise floor, preventing the
UE signals from contributing to sensing and thereby nullifying the
HRF gain. To illustrate this effect for a \emph{representative
channel realization}, we plot the HRF Pareto boundary
for different ADC resolutions in Figs.~\ref{fig:results_bit_bound_1}
and~\ref{fig:results_bit_bound_2}. For this experiment, we set $K=1$
and $P_{k}^{\mathrm{u}}=10$~dBm.

Fig.~\ref{fig:results_bit_bound_1} shows the Pareto
boundary obtained from the \emph{sensing–centric
problem}~\eqref{eq:P0_2} for different ADC resolutions.
For each $b$, we first determine the largest feasible per–UE rate
by solving~\eqref{eq:P0_2} without the CRB constraint in~\eqref{eq:p1_1_c_2},
denoted by $\Gamma_{k,\max}^{(b)}$, and then sweep $\Gamma_{k}$
from $0.1$ to $\Gamma_{k,\max}^{(b)}$. UE resources are allowed
to participate in sensing only when~\eqref{eq:bit_bound} is satisfied.
In this scenario, the condition holds for $b\geq6$ and for the ideal
ADC. Consequently, only in this regime, the CRB varies with $\Gamma_{k}$,
since the optimizer can allocate UE power toward sensing. For $b<6$,
the constraint~\eqref{eq:bit_bound} is violated, so UE resources
are excluded from sensing and the CRB becomes independent of $\Gamma_{k}$,
collapsing the Pareto curve to a constant determined solely by the
BS transmit power $P_{\mathrm{BS}}$. This behavior indicates that
HRF gains are not observable when $b<6$ under the considered channel
realization.

Fig.~\ref{fig:results_bit_bound_2} depicts the
corresponding behavior for the \emph{communication–centric
design} in~\eqref{eq:P1_1}. For each $b$, we
compute the smallest feasible CRB by solving~\eqref{eq:P0_2} without
the rate constraint in~\eqref{eq:p0_1_c_4}, denoted by $\mu_{\min}^{(b)}$,
and the largest feasible CRB by solving~\eqref{eq:P1_1} without
the CRB constraint in~\eqref{eq:p1_1_c_2}, denoted by $\mu_{\max}^{(b)}$.
The CRB constraint $\mu$ is then swept from $\mu_{\min}^{(b)}$ to
$\mu_{\max}^{(b)}$. When $b<6$,~\eqref{eq:bit_bound} is not satisfied,
and UE resources cannot support sensing, leading to a flat Pareto
boundary across $\mu$. In contrast, for $b\geq6$ and for the ideal
ADC, the maximum achievable sum spectral efficiency (SE) varies with
$\mu$, since the optimizer can trade UE resources between sensing
and communication. Together, Figs.~\ref{fig:results_bit_bound_1}
and~\ref{fig:results_bit_bound_2} demonstrate that the ADC DR governs
whether HRF trade-offs are realizable for a given channel and reveal
the CRB–rate Pareto boundary once the DR requirement is met.\setcounter{remark}{0}

\subsection{CRB-rate Trade-off with $N_{\mathrm{r}}^{\mathrm{bs}}$\label{subsec:results_N_r_bs}}

\begin{figure*}
\begin{minipage}[t]{0.30\textwidth}
\begin{center}
\includegraphics[width=1\linewidth,height=4.5cm]{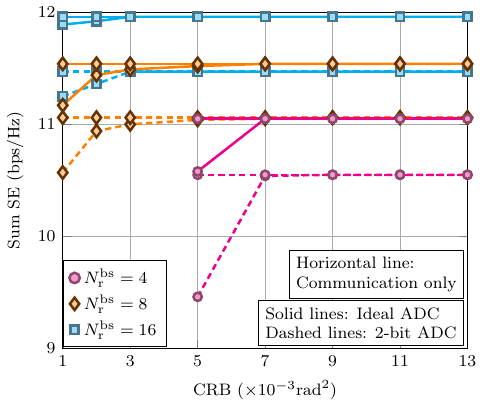}\vspace{-0.17in}
\par\end{center}
\caption{CRB-rate trade-off computed using optimization problem in \eqref{eq:P1_1}.}
\label{fig:crb_vs_rate_2}
\end{minipage}\hfill{}%
\begin{minipage}[t]{0.32\textwidth}
\begin{center}
\includegraphics[width=1\linewidth,height=4.5cm]{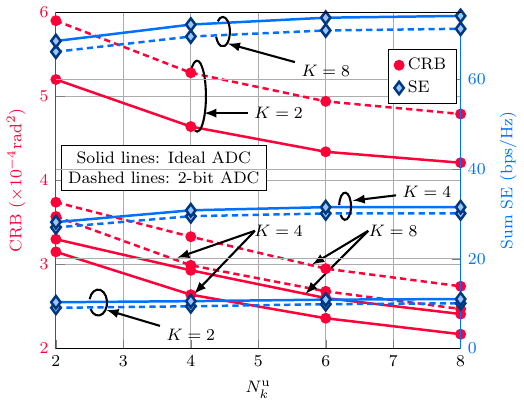}\vspace{-0.17in}
\par\end{center}
\caption{CRB and SE varying the number of transmit antennas at the UEs.}
\label{fig:CRB_SE_vs_Nu_k}
\end{minipage}\hfill{}%
\begin{minipage}[t]{0.32\textwidth}
\begin{center}
\includegraphics[width=1\linewidth,height=4.5cm]{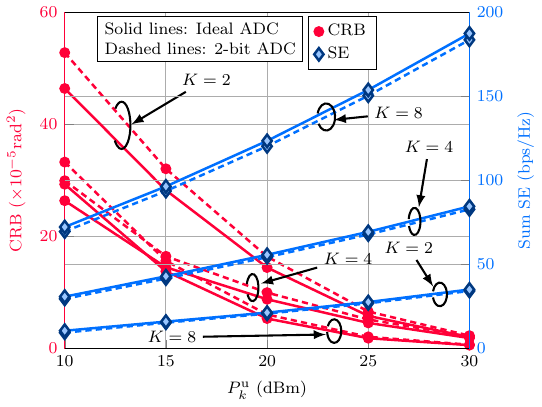}\vspace{-0.17in}
\par\end{center}
\caption{CRB and SE varying the transmit power at the UEs.}
\label{fig:CRB_SE_vs_Pu}
\end{minipage}
\vspace{-0.2in}
\end{figure*}
We next examine the impact of the number of BS receive
antennas on the CRB-rate trade-off by comparing ideal and $2$-bit
ADCs for $N_{\mathrm{r}}^{\mathrm{bs}}\in\{4,8,16\}$ in Figs.~\ref{fig:crb_vs_rate_1}
and~\ref{fig:crb_vs_rate_2}. Fig.~\ref{fig:crb_vs_rate_1} shows
the sensing-centric trade-off obtained from~\eqref{eq:P0_2}, where
$\Gamma_{k}=\Gamma$ is varied for all $k\in\mathcal{K}$. Increasing
$N_{\mathrm{r}}^{\mathrm{bs}}$ reduces the minimum achievable CRB,
but with diminishing returns beyond a certain point. For $\Gamma=1$~bps/Hz,
the $2$-bit CRB drops by about $85\%$ when $N_{\mathrm{r}}^{\mathrm{bs}}$
increases from $4$ to $8$, but by only $3.6\%$ when it increases
from $8$ to $16$, reflecting limitations imposed by transmit power
and array geometry. We also plot the \emph{sensing-only
benchmark} obtained from~\eqref{eq:P0_2} without
the rate constraint~\eqref{eq:p0_1_c_4}, which quantifies the CRB
penalty due to joint sensing–communication operation. As $\Gamma$
increases, the CRB rises sharply since more power and spatial degrees
of freedom are consumed by rate constraints. The gap between ideal
and $2$-bit ADCs remains nearly constant across antenna configurations,
corresponding to a CRB loss of approximately $13.5\%$ at $\Gamma=1$~bps/Hz.

Fig.~\ref{fig:crb_vs_rate_2} shows the communication–centric
trade-off from~\eqref{eq:P1_1}. For $N_{\mathrm{r}}^{\mathrm{bs}}=4$,
CRB values in $[1\times10^{-3},5\times10^{-3}]$~rad$^{2}$ are infeasible,
whereas they become feasible when $N_{\mathrm{r}}^{\mathrm{bs}}=8$.
Although increasing $N_{\mathrm{r}}^{\mathrm{bs}}$ improves the achievable
sum SE, the gain saturates beyond a certain point. For ${\rm CRB}=7\times10^{-3}$~rad$^{2}$,
the $2$-bit sum SE increases by $16.5\%$ when $N_{\mathrm{r}}^{\mathrm{bs}}$
in increased from $4$ to $8$, but by only $4\%$ from $8$ to $16$.
The nearly constant separation between the ideal and $2$-bit curves
indicates a comparable quantization-induced SNR loss across antenna
settings. We also plot the \emph{communication-only
benchmark} obtained from~\eqref{eq:P1_1} without
the CRB constraint~\eqref{eq:p1_1_c_2}, which shows that once the
BS alone can satisfy the sensing constraint, additional relaxation
of the CRB no longer increases the rate, which occurs around ${\rm CRB}=7\times10^{-3}$,
$5\times10^{-3}$, and $3\times10^{-3}$~rad$^{2}$ for $N_{\mathrm{r}}^{\mathrm{bs}}=4,8,16$,
respectively.\begin{remark}In Figs.~\ref{fig:CRB_SE_vs_Nu_k}--\ref{fig:CRB_SE_vs_b},
the left y-axis reports the CRB obtained from~\eqref{eq:P0_2} for
$\Gamma_{k}=\Gamma\ \forall k\in\mathcal{K}$, while the right y-axis
shows the sum SE obtained from~\eqref{eq:P1_1} for a given $\mu$.
The two curves therefore correspond to solutions of different optimization
problems.\end{remark}

\subsection{HRF performance with varying $N_{k}^{\mathrm{u}}$\label{subsec:results_N_u_k}}

Fig.~\ref{fig:CRB_SE_vs_Nu_k} shows the impact of
the UE antenna count $N_{k}^{\mathrm{u}}$ and the number of UEs $K$
on HRF sensing and communication performance. The CRB is evaluated
at $\Gamma=1$~bps/Hz and $P_{k}^{\mathrm{u}}=10$~dBm. Increasing
$N_{k}^{\mathrm{u}}$ consistently reduces the CRB, since the additional
spatial degrees of freedom at the UE strengthen the effective UL sensing
channels and increase the Fisher information. For a fixed $N_{k}^{\mathrm{u}}$,
the CRB improves when $K$ \emph{increases}
from $2$ to $4$, but \emph{degrades}
at $K=8$. This reflects a trade-off between the total number of UL
observations and the per-UE resource allocation. While a larger $K$
increases the aggregate sensing data, the fixed UL bandwidth implies
that each UE occupies $64/K$ subcarriers (i.e., $32$, $16$, and
$8$ for $K\in\{2,4,8\}$), which makes satisfying the per-UE SE constraint
increasingly power-intensive. As a result, at large $K$ the optimizer
must devote more power to communication, leaving fewer resources for
sensing and increasing the CRB. This demonstrates that, under fixed
power and bandwidth, increasing the number of users does not always
improve sensing performance.

The sum SE is evaluated at $\mu=7\times10^{-4}$~rad$^{2}$.
It increases with $N_{k}^{\mathrm{u}}$, but with diminishing returns
due to the fixed power budget. Moreover, larger $N_{k}^{\mathrm{u}}$
allows the CRB constraint to be met with less sensing-oriented power,
enabling more power to be allocated to communication. Once the CRB
constraint becomes inactive, further gains stem primarily from moderate
SNR improvements. Finally, the sum SE increases strongly with $K$,
since more users contribute to the total throughput despite the reduced
subcarrier allocation per UE.

\subsection{HRF performance with varying $P_{k}^{\mathrm{u}}$\label{subsec:results_P_u_k}}

\begin{figure*}
\centering
\begin{minipage}[t]{0.32\textwidth}
    \includegraphics[width=\linewidth,height=4.1cm]{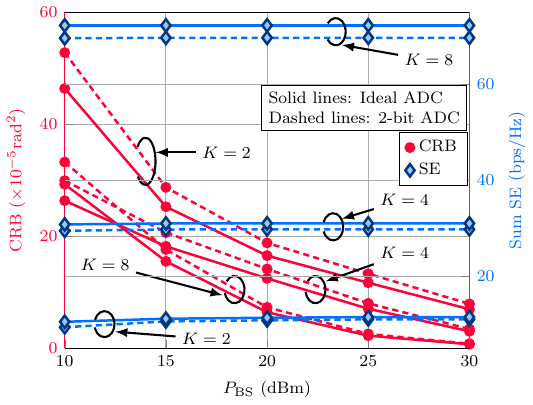}\vspace{-0.1in}
    \caption{CRB and SE varying the transmit power at the BS.}
    \label{fig:CRB_SE_vs_Pbs}
\end{minipage}\hfill
\begin{minipage}[t]{0.32\textwidth}
    \includegraphics[width=\linewidth,height=4.1cm]{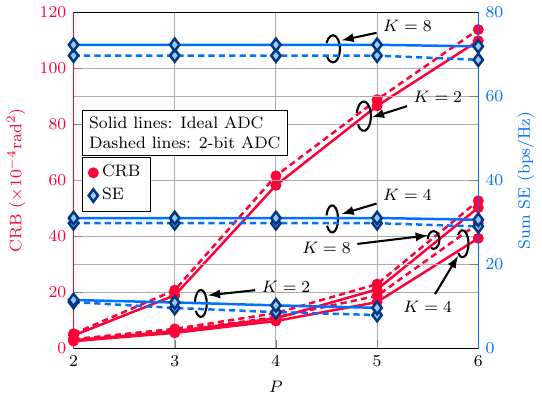}\vspace{-0.1in}
    \caption{CRB and SE varying the number of targets in the scene.}
    \label{fig:CRB_SE_vs_P}
\end{minipage}\hfill
\begin{minipage}[t]{0.32\textwidth}
    \includegraphics[width=\linewidth,height=4.1cm]{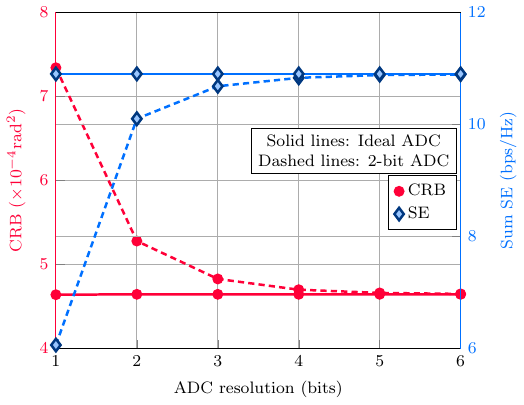}\vspace{-0.1in}
    \caption{CRB and SE versus varying ADC resolution.}
    \label{fig:CRB_SE_vs_b}
\end{minipage}
\vspace{-0.2in}
\end{figure*}

Fig.~\ref{fig:CRB_SE_vs_Pu} shows the impact of
the UE transmit power $P_{k}^{\mathrm{u}}$ for different numbers
of UEs, where $P_{k}^{\mathrm{u}}=P^{\mathrm{u}}\ \forall k\in\mathcal{K}$
and $\Gamma=1$~bps/Hz. The CRB decreases with $P^{\mathrm{u}}$
due to the improved received SNR, while the nearly constant gap between
ideal and $2$-bit ADCs reflects a resolution-dependent effective
SNR loss. Except at $P^{\mathrm{u}}=10$~dBm, the CRB decreases with
$K$, since larger $K$ provides more UL observations and, at sufficiently
high $P^{\mathrm{u}}$, the rate constraint in~\eqref{eq:p0_1_c_4}
becomes less restrictive, allowing more power to be allocated to sensing.
This effect is power dependent. Increasing $K$ from $2$ to $4$
reduces the $2$-bit CRB by about $43\%$ at $P^{\mathrm{u}}=10$~dBm
but only by $7.8\%$ at $30$~dBm, indicating diminishing returns
once the communication constraint is weak. From $K=4$ to $8$, the
CRB increases at $10$~dBm but decreases by $5.5\%$ at $15$~dBm
and by $68\%$ at $30$~dBm, since higher transmit power allows the
rate constraint to be met while exploiting the additional sensing
samples. For $\mu=7\times10^{-4}$~rad$^{2}$, the sum SE increases
monotonically with $P^{\mathrm{u}}$ for all $K$, driven by the corresponding
SNR gains.

\subsection{HRF performance with varying $P_{\mathrm{BS}}$\label{subsec:results_P_bs}}

Fig.$~\ref{fig:CRB_SE_vs_Pbs}$ shows the dependence
of the CRB and sum SE on the BS transmit power $P_{\mathrm{BS}}$
for different numbers of UEs, evaluated at $\Gamma=1$~bps/Hz and
$\mu=7\times10^{-4}$~rad$^{2}$ with $P_{k}^{\mathrm{u}}=10$~dBm.
For both ideal and $2$-bit ADCs, the CRB decreases as $P_{\mathrm{BS}}$
increases from $10$ to $30$~dBm due to the improved sensing SNR.
Except at $P_{\mathrm{BS}}=10$~dBm, the CRB also decreases with
$K$, since a larger number of UEs provides more UL observations.
This trend dominates for $P_{\mathrm{BS}}\ge15$~dBm. In contrast,
the sum SE is nearly saturated with respect to $P_{\mathrm{BS}}$,
with only a modest increase from $10$ to $15$~dBm. This is because
at low $P_{\mathrm{BS}}$ the UEs must allocate part of their power
to satisfy the CRB constraint in~\eqref{eq:p1_1_c_2}, whereas for
$P_{\mathrm{BS}}\ge15$~dBm the BS alone meets the sensing requirement,
allowing the UEs to transmit at full power. Since the UE power is
fixed, further increases in $P_{\mathrm{BS}}$ do not improve the
sum SE, which instead scales primarily with $K$ as more users contribute
to the total throughput.

\subsection{HRF performance with varying $P$\label{subsec:results_P}}

Fig.~\ref{fig:CRB_SE_vs_P} shows the dependence
of the CRB and sum SE on the number of targets $P$ for different
numbers of UEs $K$, evaluated at $\Gamma=1$~bps/Hz, $\mu=7\times10^{-3}$~rad$^{2}$,
and $P_{k}^{\mathrm{u}}=10$~dBm. The CRB increases monotonically
with $P$ since more unknown target parameters must be estimated with
fixed sensing power and observations, reducing the information available
per target. For all $P$, increasing $K$ from $2$ to $4$ lowers
the CRB, whereas increasing it to $8$ slightly raises it due to the
tighter per-UE SE constraint and reduced bandwidth per UE, which force
more power to be allocated to communication rather than sensing, an
effect that becomes more pronounced as $P$ grows. The sum SE decreases
with $P$ for $K=2$ because more UE resources are diverted to satisfy
the CRB constraint, while for $K=4$ and $K=8$ it remains nearly
constant over $P=2$–$5$ before dropping at larger $P$ (here $P=6$),
reflecting that the higher total transmit power and larger number
of UL observations can accommodate moderate sensing loads but eventually
become insufficient as the number of targets increases.

\subsection{HRF performance with varying ADC resolution\label{subsec:results_b}}

Fig.~\ref{fig:CRB_SE_vs_b} illustrates the dependence
of the CRB and sum SE on the ADC resolution $b$, evaluated at $\Gamma=1$~bps/Hz,
$\mu=8\times10^{-4}$~rad$^{2}$, and $P_{k}^{\mathrm{u}}=10$~dBm.
As $b$ increases, both the CRB and sum SE improve due to reduced
quantization distortion and higher effective SNR, with most of the
gain realized between $b=1$ and $b=6$. For $b\ge5$ and $6$, the
performance closely approaches the ideal-ADC benchmark. However, since
the LSB determines the ADC DR via~\eqref{eq:bit_bound}, low-resolution
ADCs, while energy efficient, may fail to resolve weak reflections
in the presence of strong DPs. This highlights a fundamental trade-off
between energy efficiency and the DR requirements of HRF.

\section{Conclusion and Future works\label{sec:results_conclusion}}

This work characterized how the ADC DR at the BS
fundamentally constrains HRF performance. We developed a unified quantized
HRF framework by deriving Bussgang-based expressions for an achievable
UL rate and a tractable CRB, and by linking ADC resolution to system
operation through a detectability condition that guarantees resolvability
of the weakest target reflections in the presence of the strongest
DP. Based on these results, we formulated complementary sensing-centric
and communication-centric optimization problems that enable a consistent
comparison between ideal and finite-resolution receivers. Numerical
results revealed that ADC DR is the dominant performance bottleneck
and, when the DR requirement is satisfied, how the CRB–rate trade-off
scales with the number of UE and BS antennas, number of UEs and targets,
transmit powers, and ADC resolution, yielding concrete design guidelines
for HRF systems.

\appendices{}

\section{Proof of Proposition 1\label{sec:prop_1_proof}}

For a symmetric input PDF and symmetric quantizer
\cite{mezghani_capacity_nodate}, the properties, $\mathbb{E}\big[q_{i,n}^{(\ell,\zeta)}\big]=0,$
and $\mathbb{E}\big[r_{i,n}^{(\ell,\zeta)}q_{i,n}^{(\ell,\zeta)}\big]=0$
hold for all $i\in\left\{ 1,\ldots,N_{\mathrm{r}}^{\mathrm{bs}}\right\} $,
$n\in\left\{ 0,\ldots,N_{\mathrm{c}}-1\right\} $, $\ell\in\left\{ 0,\ldots,L-1\right\} $,
and $\zeta\in\left\{ R,I\right\} $, with $R$ and $I$ denoting the
real and imaginary parts of the complex variable, where $r_{i,n}^{(\ell,\zeta)}=y_{i,n}^{(\ell,\zeta)}+q_{i,n}^{(\ell,\zeta)}$,
$r_{i,n}^{(\ell,\zeta)}$ is the real or imaginary part of the $n^{\mathrm{th}}$
received time sample of the $\ell^{\mathrm{th}}$ OFDM symbol on the
$i^{\mathrm{th}}$ antenna and the distortion factor for this sample
is defined as $\gamma_{i,n}^{(\ell,\zeta)}=\frac{\mathbb{E}\big[\big(q_{i,n}^{(\ell,\zeta)}\big)^{2}\big]}{r_{y_{i,n}^{(\ell,\zeta)}y_{i,n}^{(\ell,\zeta)}}},\:r_{y_{i,n}^{(\ell,\zeta)}y_{i,n}^{(\ell,\zeta)}}\triangleq\mathbb{E}\big[\big(y_{i,n}^{(\ell,\zeta)}\big)^{2}\big].$
For a Gaussian distributed\footnote{By the argument in Footnote 1, we model the quantizer
inputs as Gaussian.} $y_{i,c}^{(\ell,\zeta)}$ and identical ADCs across
all RF chains of the BS, $\gamma_{i,n}^{(\ell,\zeta)}=\gamma$ $\forall i\in\left\{ 1,\ldots,N_{\mathrm{r}}^{\mathrm{bs}}\right\} $,
$n\in\left\{ 0,\ldots,N_{\mathrm{c}}-1\right\} $, $\ell\in\left\{ 0,\ldots,L-1\right\} $,
and $\zeta\in\left\{ R,I\right\} $. Moreover, the real and imaginary
parts of $y_{i,n}^{(\ell)}$ and $q_{i,n}^{(\ell)}$ are uncorrelated.

The covariance matrix, $\mathbf{R}_{\mathbf{w}_{\mathrm{q}}^{(\ell)}\mathbf{w}_{\mathrm{q}}^{(\ell)}}[\tau]$
can be written as 
\begin{flalign*}
 & \mathbf{R}_{\mathbf{w}_{\mathrm{q}}^{(\ell)}\mathbf{w}_{\mathrm{q}}^{(\ell)}}[\tau]=\mathbb{E}\big[\big(\mathbf{r}_{n}^{(\ell)}-\eta\mathbf{y}_{n}^{(\ell)}\big)\big(\mathbf{r}_{n-\tau}^{(\ell)}-\eta\mathbf{y}_{n-\tau}^{(\ell)}\big)\big]^{\mathrm{H}},\\
 & =\mathbf{R}_{\mathbf{r}^{(\ell)}\mathbf{r}^{(\ell)}}[\tau]\!\!-\!\!\eta\mathbf{R}_{\mathbf{y}^{(\ell)}\mathbf{r}^{(\ell)}}^{\mathrm{H}}[-\tau]\!\!-\!\!\eta\mathbf{R}_{\mathbf{y}^{(\ell)}\mathbf{r}^{(\ell)}}[\tau]\!+\!\eta^{2}\mathbf{R}_{\mathbf{y}^{(\ell)}\mathbf{y}^{(\ell)}}[\tau].
\end{flalign*}
In order to derive $\mathbf{R}_{\mathbf{w}_{\mathrm{q}}^{(\ell)}\mathbf{w}_{\mathrm{q}}^{(\ell)}}[\tau]$,
$\mathbf{R}_{\mathbf{r}^{(\ell)}\mathbf{r}^{(\ell)}}[\tau]$ and $\mathbf{R}_{\mathbf{y}^{(\ell)}\mathbf{r}^{(\ell)}}[\tau]$
have to be computed. $\mathbf{R}_{\mathbf{r}^{(\ell)}\mathbf{r}^{(\ell)}}[\tau]$
is given as follows
\begin{flalign*}
 & \mathbf{R}_{\mathbf{r}^{(\ell)}\mathbf{r}^{(\ell)}}[\tau]\!\!=\!\!\mathbb{E}\big[\mathbf{(y}^{(\ell)}[n]\!\!+\!\!\mathbf{q}^{(\ell)}[n])\mathbf{(y}^{(\ell)}[n-\tau]\!\!+\!\!\mathbf{q}^{(\ell)}[n-\tau])^{\mathrm{H}}\big],\\
 & =\!\!\mathbf{R}_{\mathbf{y}^{(\ell)}\mathbf{y}^{(\ell)}}[\tau]\!\!+\!\!\mathbf{R}_{\mathbf{y}^{(\ell)}\mathbf{q}^{(\ell)}}[\tau]\!\!+\!\!\mathbf{R}_{\mathbf{y}^{(\ell)}\mathbf{q}^{(\ell)}}^{\mathrm{H}}[-\tau]\!\!+\!\!\mathbf{R}_{\mathbf{q}^{(\ell)}\mathbf{q}^{(\ell)}}[\tau].\!\!\!\!
\end{flalign*}
Therefore, to compute $\mathbf{R}_{\mathbf{r}^{(\ell)}\mathbf{r}^{(\ell)}}[\tau]$,
$\mathbf{R}_{\mathbf{y}^{(\ell)}\mathbf{q}^{(\ell)}}[\tau]$ and $\mathbf{R}_{\mathbf{q}^{(\ell)}\mathbf{q}^{(\ell)}}[\tau]$
have to be evaluated. The derivations for both the covariance matrices
for all the cases are given as follows.
\begin{flalign}
 & r_{q_{i}^{(\ell)}q_{i}^{(\ell)}}[0]=\mathbb{E}\big[q_{i,n}^{(\ell)}q_{i,n}^{(\ell)*}\big]=\mathbb{E}\big[\big(q_{i,n}^{(\ell,R)}\big)^{2}+\mathbb{E}\big[\big(q_{i,n}^{(\ell,I)}\big)^{2}\big],\nonumber \\
 & =\gamma r_{y_{i,n}^{(\ell,R)}y_{i,n}^{(\ell,R)}}+\gamma r_{y_{i,n}^{(\ell,I)}y_{i,n}^{(\ell,I)}}=\gamma r_{y_{i}^{(\ell)}y_{i}^{(\ell)}}[0].\label{eq:r_qi_qi}
\end{flalign}
\begin{flalign}
 & r_{y_{i}^{(\ell)}q_{i}^{(\ell)}}[0]=\mathbb{E}\big[y_{i,n}^{(\ell)}q_{i,n}^{(\ell)*}\big]=\mathbb{E}\big[\big(r_{i,n}^{(\ell)}-q_{i,n}^{(\ell)}\big)q_{i,n}^{(\ell)*}\big],\nonumber \\
 & =\mathbb{E}\big[r_{i,n}^{(\ell)}q_{i,n}^{(\ell)*}\big]-\mathbb{E}\big[\big(q_{i,n}^{(\ell)}\big){}^{2}\big]=-\gamma r_{y_{i}^{(\ell)}y_{i}^{(\ell)}}[0].\label{eq:r_yi_qi}
\end{flalign}
\begin{flalign}
 & r_{y_{i}^{(\ell)}q_{i}^{(\ell)}}[\tau]=\mathbb{E}\big[y_{i,n}^{(\ell)}q_{i,n-\tau}^{(\ell)*}\big]=\mathbb{E}_{y_{i,n-\tau}^{(\ell)}}\big[\mathbb{E}\big[y_{i,n}^{(\ell)}q_{i,n-\tau}^{(\ell)*}|y_{i,n-\tau}^{(\ell)}\big]\big],\nonumber \\
 & =\mathbb{E}_{y_{i,n-\tau}^{(\ell)}}\big[\mathbb{E}\big[y_{i,n}^{(\ell)}|y_{i,n-\tau}^{(\ell)}\big]\mathbb{E}\big[q_{i,n-\tau}^{(\ell)*}|y_{i,n-\tau}^{(\ell)}\big]\big],\nonumber \\
 & \approx\mathbb{E}_{y_{i,n-\tau}^{(\ell)}}\bigg[\tfrac{r_{y_{i}^{(\ell)}y_{i}^{(\ell)}}[\tau]}{r_{y_{i}^{(\ell)}y_{i}^{(\ell)}}[0]}y_{i,n-\tau}^{(\ell)}\mathbb{E}\big[q_{i,n-\tau}^{(\ell)*}|y_{i,n-\tau}^{(\ell)}\big]\bigg],\nonumber \\
 & =r_{y_{i}^{(\ell)}y_{i}^{(\ell)}}[\tau]r_{y_{i}^{(\ell)}y_{i}^{(\ell)}}^{-1}[0]\mathbb{E}\big[y_{i,n-\tau}^{(\ell)}q_{i,n-\tau}^{(\ell)*}\big],\nonumber \\
 & =r_{y_{i}^{(\ell)}y_{i}^{(\ell)}}[\tau]r_{y_{i}^{(\ell)}y_{i}^{(\ell)}}^{-1}[0]\big(-\gamma r_{y_{i}^{(\ell)}y_{i}^{(\ell)}}[0]\big),\nonumber \\
 & =-\gamma r_{y_{i}^{(\ell)}y_{i}^{(\ell)}}[\tau].\label{eq:r_yi_qi_tau}
\end{flalign}
We use the approximation given in~\cite{mezghani_capacity_nodate}
to obtain \eqref{eq:r_yi_qi_tau}. Following a similar methodology
and approximation, $r_{y_{i}^{(\ell)}q_{j}^{(\ell)}}[\tau]$, $r_{q_{i}^{(\ell)}q_{i}^{(\ell)}}[\tau]$,
and $r_{q_{i}^{(\ell)}q_{j}^{(\ell)}}[\tau]$, can be obtained, which
are given as follows
\begin{flalign}
r_{y_{i}^{(\ell)}q_{j}^{(\ell)}}[\tau] & =-\gamma r_{y_{i}^{(\ell)}y_{j}^{(\ell)}}[\tau],\label{eq:r_yi_qj_tau}\\
r_{q_{i}^{(\ell)}q_{i}^{(\ell)}}[\tau] & =\gamma^{2}r_{y_{i}^{(\ell)}y_{i}^{(\ell)}}[\tau],\label{eq:r_qi_qi_tau}\\
r_{q_{i}^{(\ell)}q_{j}^{(\ell)}}[\tau] & =\gamma^{2}r_{y_{i}^{(\ell)}y_{j}^{(\ell)}}[\tau].\label{eq:r_qi_qj_tau}
\end{flalign}
We omit the detailed derivation of all cases due to the space limitation.
Thus, from \eqref{eq:r_yi_qi}, \eqref{eq:r_yi_qi_tau}, \eqref{eq:r_yi_qj_tau},
\eqref{eq:r_qi_qi}, \eqref{eq:r_qi_qi_tau}, and \eqref{eq:r_qi_qj_tau},
$\mathbf{R}_{\mathbf{y}^{(\ell)}\mathbf{q}^{(\ell)}}[\tau]$, and
$\mathbf{R}_{\mathbf{q}^{(\ell)}\mathbf{q}^{(\ell)}}$can be written
as follows
\begin{equation}
\mathbf{R}_{\mathbf{y}^{(\ell)}\mathbf{q}^{(\ell)}}[\tau]=\mathbf{R}_{\mathbf{y}^{(\ell)}\mathbf{q}^{(\ell)}}^{\mathrm{H}}[-\tau]=-\gamma\mathbf{R}_{\mathbf{y^{(\ell)}\mathbf{y}^{(\ell)}}}[\tau].\label{eq:R_y_q}
\end{equation}
\begin{multline}
\mathbf{R}_{\mathbf{q}^{(\ell)}\mathbf{q}^{(\ell)}}[\tau]=\begin{cases}
\gamma^{2}\mathbf{R}_{\mathbf{y}^{(\ell)}\mathbf{y}^{(\ell)}}[\tau], & \tau\neq0\\
\gamma\mathrm{diag}\big(\mathbf{R}_{\mathbf{y}^{(\ell)}\mathbf{y}^{(\ell)}}[0]\big)\\
+\mathrm{\gamma^{2}nondiag}\big(\mathrm{\mathbf{R}_{\mathbf{y}^{(\ell)}\mathbf{y}^{(\ell)}}[0]\big)}. & \tau=0
\end{cases}\label{eq:R_q_q}
\end{multline}
Using \eqref{eq:R_y_q} and \eqref{eq:R_q_q}, for $\tau=0$ and $\tau\neq0$,
$\mathbf{R}_{\mathbf{r}^{(\ell)}\mathbf{r}^{(\ell)}}[\tau]$ can be
written as
\begin{flalign}
 & \mathbf{R}_{\mathbf{r}^{(\ell)}\mathbf{r}^{(\ell)}}[0]=\mathbf{R}_{\mathbf{y}^{(\ell)}\mathbf{y}^{(\ell)}}[0]+\mathbf{R}_{\mathbf{y}^{(\ell)}\mathbf{q}^{(\ell)}}[0]+\mathbf{R}_{\mathbf{y}^{(\ell)}\mathbf{q}^{(\ell)}}^{\mathrm{H}}[0]\nonumber \\
 & \qquad\qquad\qquad\;+\mathbf{R}_{\mathbf{q}^{(\ell)}\mathbf{q}^{(\ell)}}[0],\nonumber \\
 & =(1-\gamma)\big(\mathrm{diag}\big(\mathbf{R}_{\mathbf{y}^{(\ell)}\mathbf{y}^{(\ell)}}[0]\big)\nonumber \\
 & \quad+(1-\gamma)\mathrm{nondiag}\big(\mathbf{R}_{\mathbf{y}^{(\ell)}\mathbf{y}^{(\ell)}}[0]\big)\big).\label{eq:R_r_r}
\end{flalign}
\begin{flalign}
 & \mathbf{R}_{\mathbf{r}^{(\ell)}\mathbf{r}^{(\ell)}}[\tau]=\mathbf{R}_{\mathbf{y}^{(\ell)}\mathbf{y}^{(\ell)}}[\tau]+\mathbf{R}_{\mathbf{y}^{(\ell)}\mathbf{q}^{(\ell)}}[\tau]+\mathbf{R}_{\mathbf{y}^{(\ell)}\mathbf{q}^{(\ell)}}^{\mathrm{H}}[\tau]\nonumber \\
 & \qquad\qquad\qquad\;+\mathbf{R}_{\mathbf{q}^{(\ell)}\mathbf{q}^{(\ell)}}[\tau],\nonumber \\
 & =(1-\gamma)^{2}\mathbf{R}_{\mathbf{y}^{(\ell)}\mathbf{y}^{(\ell)}}[\tau].\label{eq:R_r_r_tau}
\end{flalign}
Hereafter, we compute $\mathbf{R}_{\mathbf{y}^{(\ell)}\mathbf{r}^{(\ell)}}[\tau]$.
\begin{flalign}
 & \mathbf{R}_{\mathbf{y}^{(\ell)}\mathbf{r}^{(\ell)}}[\tau]=\mathbb{E}\big[\mathbf{y}^{(\ell)}[n]\mathbf{(r}^{(\ell)}[n-\tau])^{\mathrm{H}}\big],\nonumber \\
 & \qquad\qquad\;\:\:\,=\left(1-\gamma\right)\mathbf{R}_{\mathbf{y}^{(\ell)}\mathbf{y}^{(\ell)}}[\tau].\label{eq:R_y_r_tau}
\end{flalign}
Using \eqref{eq:R_r_r}, \eqref{eq:R_r_r_tau}, and \eqref{eq:R_y_r_tau},
for $\tau=0$ and $\tau\neq0$, $\mathbf{R}_{\mathbf{w}_{\mathrm{q}}^{(\ell)}\mathbf{w}_{\mathrm{q}}^{(\ell)}}[\tau]$
is given as
\begin{flalign}
 & \mathbf{R}_{\mathbf{w}_{\mathrm{q}}^{(\ell)}\mathbf{w}_{\mathrm{q}}^{(\ell)}}[0]=\mathbf{R}_{\mathbf{r}^{(\ell)}\mathbf{r}^{(\ell)}}[0]-\eta\mathbf{R}_{\mathbf{y}^{(\ell)}\mathbf{r}^{(\ell)}}^{\mathrm{H}}[0]-\eta\mathbf{R}_{\mathbf{y}^{(\ell)}\mathbf{r}^{(\ell)}}[0]\nonumber \\
 & \qquad\qquad\qquad\;+\eta^{2}\mathbf{R}_{\mathbf{y}^{(\ell)}\mathbf{y}^{(\ell)}}[0],\nonumber \\
 & =\eta(1-\eta)\mathrm{diag}\left(\mathbf{R}_{\mathbf{y}^{(\ell)}\mathbf{y}^{(\ell)}}[0]\right).\label{eq:quant_noise_cov_zero}
\end{flalign}
\begin{flalign}
 & \mathbf{R}_{\mathbf{w}_{\mathrm{q}}^{(\ell)}\mathbf{w}_{\mathrm{q}}^{(\ell)}}[\tau]=\mathbf{R}_{\mathbf{r}^{(\ell)}\mathbf{r}^{(\ell)}}[\tau]-\eta\mathbf{R}_{\mathbf{y}^{(\ell)}\mathbf{r}^{(\ell)}}^{\mathrm{H}}[\tau]-\eta\mathbf{R}_{\mathbf{y}^{(\ell)}\mathbf{r}^{(\ell)}}[\tau]\nonumber \\
 & \qquad\qquad\qquad\;+\eta^{2}\mathbf{R}_{\mathbf{y}^{(\ell)}\mathbf{y}^{(\ell)}}[\tau],\nonumber \\
 & \!\!=\!\!2(1-\gamma)^{2}\mathbf{R}_{\mathbf{y}^{(\ell)}\mathbf{y}^{(\ell)}}[\tau]-2(1-\gamma)^{2}\mathbf{R}_{\mathbf{y}^{(\ell)}\mathbf{y}^{(\ell)}}[\tau]=\mathbf{0}.\!\!\!\label{eq:quant_noise_cov_delayed}
\end{flalign}

The frequency domain covariance of the quantization
distortion, $\mathbf{R}_{\tilde{\mathbf{w}}_{\mathrm{q}}^{(\ell)}\tilde{\mathbf{w}}_{\mathrm{q}}^{(\ell)}}[m]$
is computed as follows~\cite{Emil_rate_analysis}
\begin{flalign}
 & \mathbf{R}_{\tilde{\mathbf{w}}_{\mathrm{q}}^{(\ell)}\tilde{\mathbf{w}}_{\mathrm{q}}^{(\ell)}}[m]=\mathbb{E}\big[\tilde{\mathbf{w}}_{\mathrm{q}}^{(\ell)}[m]\big(\tilde{\mathbf{w}}_{\mathrm{q}}^{(\ell)}[m]\big)^{\mathrm{H}}\big],\nonumber \\
 & =\frac{1}{N_{\mathrm{c}}}\sum_{n=0}^{N_{\mathrm{c}}-1}\sum_{n'=0}^{N_{\mathrm{c}}-1}\mathbb{E}\big[\mathbf{w}_{\mathrm{q}}^{(\ell)}[n]\big(\mathbf{w}_{\mathrm{q}}^{(\ell)}[n]\big)^{\mathrm{H}}\big]e^{-j2\pi\frac{nm}{N_{\mathrm{c}}}}e^{j2\pi\frac{n'm}{N_{\mathrm{c}}}},\nonumber \\
 & =\frac{1}{N_{\mathrm{c}}}\sum_{n=0}^{N_{\mathrm{c}}-1}\sum_{n'=0}^{N_{\mathrm{c}}-1}\mathbf{R}_{\mathbf{w}_{\mathrm{q}}^{(\ell)}\mathbf{w}_{\mathrm{q}}^{(\ell)}}[n-n']e^{-j2\pi\frac{nm}{N_{\mathrm{c}}}}e^{j2\pi\frac{n'm}{N_{\mathrm{c}}}},\nonumber \\
 & =\frac{1}{N_{\mathrm{c}}}\sum\nolimits_{\tau=-(N_{\mathrm{c}}-1)}^{N_{\mathrm{c}}-1}\big(N_{\mathrm{c}}-|\tau|\big)\mathbf{R}_{\mathbf{w}_{\mathrm{q}}^{(\ell)}\mathbf{w}_{\mathrm{q}}^{(\ell)}}[\tau]e^{-j2\pi\frac{\tau m}{N_{\mathrm{c}}}},\nonumber \\
 & =\frac{1}{N_{\mathrm{c}}}\sum\nolimits_{\tau=0}^{N_{\mathrm{c}}-1}\big(N_{\mathrm{c}}-\tau\big)\big(\mathbf{R}_{\mathbf{w}_{\mathrm{q}}^{(\ell)}\mathbf{w}_{\mathrm{q}}^{(\ell)}}[\tau]e^{-j2\pi\frac{\tau m}{N_{\mathrm{c}}}}\nonumber \\
 & +\mathbf{R}_{\mathbf{w}_{\mathrm{q}}^{(\ell)}\mathbf{w}_{\mathrm{q}}^{(\ell)}}^{*}[\tau]e^{j2\pi\frac{\tau m}{N_{\mathrm{c}}}}\big)-N_{\mathrm{c}}\mathbf{R}_{\mathbf{w}_{\mathrm{q}}^{(\ell)}\mathbf{w}_{\mathrm{q}}^{(\ell)}}[0].\label{eq:freq_dom_quant_noise_cov_1}
\end{flalign}
Substituting \eqref{eq:quant_noise_cov_zero} and \eqref{eq:quant_noise_cov_delayed}
in \eqref{eq:freq_dom_quant_noise_cov_1} gives
\begin{flalign}
 & \mathbf{R}_{\tilde{\mathbf{w}}_{\mathrm{q}}^{(\ell)}\tilde{\mathbf{w}}_{\mathrm{q}}^{(\ell)}}[m]=\frac{1}{N_{\mathrm{c}}}\big(N_{\mathrm{c}}\mathbf{R}_{\mathbf{w}_{\mathrm{q}}^{(\ell)}\mathbf{w}_{\mathrm{q}}^{(\ell)}}[0]+N_{\mathrm{c}}\mathbf{R}_{\mathbf{w}_{\mathrm{q}}^{(\ell)}\mathbf{w}_{\mathrm{q}}^{(\ell)}}^{*}[0]\nonumber \\
 & \qquad\qquad\qquad\qquad-N_{\mathrm{c}}\mathbf{R}_{\mathbf{w}_{\mathrm{q}}^{(\ell)}\mathbf{w}_{\mathrm{q}}^{(\ell)}}[0]\big),\nonumber \\
 & \qquad\qquad\quad\;\;=\eta(1-\eta)\mathrm{diag}\big(\mathbf{R}_{\mathbf{y}^{(\ell)}\mathbf{y}^{(\ell)}}[0]\big).\label{eq:freq_dom_quant_noise_cov_2}
\end{flalign}
$\mathbf{R}_{\mathbf{y}^{(\ell)}\mathbf{y}^{(\ell)}}[0]$ is computed
as follows
\begin{flalign}
 & \!\!\!\!\mathbf{R}_{\mathbf{y}^{(\ell)}\mathbf{y}^{(\ell)}}[\tau]=\mathbb{E}\big[\mathrm{\mathbf{y}}_{n}^{(\ell)}\big(\mathrm{\mathbf{y}}_{n-\tau}^{(\ell)}\big)^{\mathrm{H}}\big]\nonumber \\
 & \!\!\!\!=\frac{1}{N_{\mathrm{c}}}\sum_{m\in\mathcal{N}_{\mathrm{c}}}\sum_{m'\in\mathcal{N}_{\mathrm{c}}}\mathbb{E}\big[\mathrm{\mathbf{y}}_{m}^{(\ell)}\big(\mathrm{\mathbf{y}}_{m'}^{(\ell)}\big)^{\mathrm{H}}\big]e^{j2\pi\frac{nm}{N_{\mathrm{c}}}}e^{-j2\pi\frac{(n-\tau)m'}{N_{\mathrm{c}}}},\label{eq:R_y_y_freq_1}\\
 & \!\!\!\!=\frac{1}{N_{\mathrm{c}}}\sum\nolimits_{m'\in\mathcal{N}_{\mathrm{c}}}\mathbf{R}_{\mathrm{\tilde{\mathbf{y}}}^{(\ell)}\tilde{\mathbf{y}}^{(\ell)}}[m']e^{j2\pi\frac{\tau m}{N_{\mathrm{c}}}}.\label{eq:R_y_y_freq_2}
\end{flalign}
We obtain \eqref{eq:R_y_y_freq_2} from \eqref{eq:R_y_y_freq_1} because
of the assumption of i.i.d symbols in our system model and thus, for
$\tau=0$, we get,
\begin{equation}
\mathbf{R}_{\mathbf{y}^{(\ell)}\mathbf{y}^{(\ell)}}[0]=\frac{1}{N_{\mathrm{c}}}\sum\nolimits_{m'\in\mathcal{N}_{\mathrm{c}}}\mathbf{R}_{\mathrm{\tilde{\mathbf{y}}}^{(\ell)}\tilde{\mathbf{y}}^{(\ell)}}[m'].\label{eq:freq_dom_rx_sig_cov_1}
\end{equation}
Substituting \eqref{eq:freq_dom_rx_sig_cov_1} in \eqref{eq:freq_dom_quant_noise_cov_2},
we get 
\begin{equation}
\!\!\!\!\!\!\mathbf{R}_{\tilde{\mathbf{w}}_{\mathrm{q}}^{(\ell)}\tilde{\mathbf{w}}_{\mathrm{q}}^{(\ell)}}[m]=\tfrac{\eta(1-\eta)}{N_{\mathrm{c}}}\mathrm{diag}\bigg(\sum\nolimits_{m'\in\mathcal{N}_{\mathrm{c}}}\!\!\mathbf{R}_{\mathrm{\tilde{\mathbf{y}}}^{(\ell)}\tilde{\mathbf{y}}^{(\ell)}}[m']\bigg).
\end{equation}
\bibliographystyle{IEEEtran}
\bibliography{ref}
\end{document}